\documentclass{article}
\usepackage{graphicx}
\usepackage{cite}
\usepackage{bbm}
\usepackage{amsmath,amsthm,amssymb}
\usepackage{enumitem}
\usepackage{mathtools}
\usepackage[utf8]{inputenc}
\usepackage{tabularx}
\usepackage{xcolor}
\usepackage{url}
\usepackage{hyperref}
\usepackage{subfiles}
\usepackage{algorithm}
\usepackage{algpseudocode}
\usepackage{mathdots}
\usepackage{mathabx}
\usepackage{footmisc}

\usepackage{subfig}
\usepackage{cleveref}

\newtheorem{theorem}{Theorem}[section]
\newtheorem{lemma}[theorem]{Lemma}

\theoremstyle{definition}

\newtheorem{remark}[theorem]{Remark}

\newtheorem{question}[theorem]{Question}
\newtheorem*{question*}{Question}
\newtheorem{problem}[theorem]{Problem}
\newtheorem*{problem*}{Problem}
\newtheorem{questions}[theorem]{Questions}
\newtheorem*{questions*}{Questions}
\newtheorem*{example*}{Example}

\newtheorem*{remark*}{Remark}
\newtheorem*{remarks*}{Remarks}
\newtheorem*{definition*}{Definition}

\DeclareMathOperator{\ii}{\mathbbm{i}}
\DeclareMathOperator{\MMSE}{MMSE}
\DeclareMathOperator{\diag}{diag}
\DeclareMathOperator{\sign}{sign}

\title{Phasebook:  A Survey of Selected Open Problems in Phase Retrieval}

\author{Marc Allain\footnote{Aix Marseille Université, CNRS, Centrale Marseille, Institut Fresnel, Marseille, France, \emph{marc.allain@fresnel.fr}.} \and
Selin Aslan\footnote{Koc University, Department of Mathematics, College of Science, Sarıyer Rumeli Feneri Yolu, 34450 Sarıyer, Istanbul, Turkey, \emph{saslan@ku.edu.tr}.} \and
Wim Coene\footnote{Delft University of Technology, Applied Sciences, Imaging Physics, Mekelweg 5, 2628 CD Delft, The Netherlands, \emph{w.m.j.m.coene-1@tudelft.nl}.} \and
Sjoerd Dirksen\footnote{Utrecht University, Mathematical Institute, Budapestlaan 6, 3584 CD Utrecht, The Netherlands, \emph{s.dirksen@uu.nl}.} \and 
Jonathan Dong\footnote{Ecole Polytechnique Fédérale de Lausanne, Biomedical Imaging Group, 1015 Lausanne, Switzerland, \emph{jonathan.dong@epfl.ch}.} \and 
Julien Flamant\footnote{Universit\'{e} de Lorraine, CNRS, CRAN, F-54000 Nancy, France, \emph{julien.flamant@cnrs.fr}.} \and 
Mark Iwen\footnote{Michigan State University, Department of Mathematics, and Department of Computational Mathematics, Science and Engineering (CMSE), \emph{iwenmark@msu.edu}.} \and
Felix Krahmer\footnote{Technical University of Munich, Department of Mathematics, Boltzmannstr. 3, 85748 Garching, Germany,  \emph{felix.krahmer@tum.de}.} \and 
Tristan van Leeuwen\footnote{Utrecht University, Mathematical Institute, Budapestlaan 6, Utrecht; and Centrum Wiskunde \& Informatica, Computational Imaging, Science Park 123 Amsterdam, \emph{t.van.leeuwen@cwi.nl}.} \and 
Oleh Melnyk\footnote{Technische Universität Berlin, Institute of Mathematics, Straße des 17. Juni 136, D-10623 Berlin, Germany, \emph{oleh.melnyk@tu-berlin.de} and Ludwig-Maximilians-Universität München, Mathematical Foundation of Artificial Intelligence, Akademiestr. 7, 80799 Munich, Germany.} \and 
Andreas Menzel\footnote{Paul Scherrer Institute, Center for Photon Science, 5232 Villigen PSI, Switzerland, \emph{andreas.menzel@psi.ch}.} \and  
Allard P. Mosk\footnote{Utrecht University, Debye Institute for Nanomaterials Science, Princetonplein 1, Utrecht, \emph{a.p.mosk@uu.nl}.} \and 
Viktor Nikitin\footnote{Argonne National Laboratory, Advanced Photon Source, 9700 S Cass Ave, Lemont, IL 60439, USA, \emph{vnikitin@anl.lgov}.} \and 
Palina Salanevich\footnote{Utrecht University, Mathematical Institute, Budapestlaan 6, 3584CD Utrecht, the Netherlands, \emph{p.salanevich@uu.nl}.} \and 
Gerlind Plonka\footnote{University of Goettingen, Institute for Numerical and Applied Mathematics, Lotzestrasse 16-18, 37083 Goettingen, Germany, \emph{plonka@math.uni-goettingen.de}.} \and 
Matthias Wellershoff\footnote{University of Maryland, Department of Mathematics, 4176 Campus Drive, College Park, MD 20742, \emph{wellersm@umd.edu}.} 
}

\date{}

\begin{document}

\maketitle

\newpage

\begin{abstract}
    Phase retrieval is an inverse problem that, on one hand, is crucial in many applications across imaging and physics, and, on the other hand, leads to deep research questions in theoretical signal processing and applied harmonic analysis. This survey paper is an outcome of the recent workshop \textit{Phase Retrieval in Mathematics and Applications (PRiMA)} (held on August 5--9 2024 at the Lorentz Center in Leiden, The Netherlands) that brought together experts working on theoretical and practical aspects of the phase retrieval problem with the purpose to formulate and explore essential open problems in the field.
    
    \medskip

\end{abstract}



\section{Introduction}
\label{S:Introduction}
Phase retrieval is a fundamental problem in signal processing and imaging in which one seeks to recover a signal from magnitude-only measurements. It arises in many areas of the physical sciences --- from X-ray crystallography and astronomy to electron microscopy and quantum tomography --- whenever detectors record intensity without phase information. Because the phase information is lost, the inverse reconstruction problem becomes highly ill-posed, raising fundamental questions regarding the uniqueness and stability of solutions and efficient algorithms to find them. 
\subsection{Purpose of the Survey}
Despite these inherent challenges, the field of phase retrieval has witnessed significant advances, driven by fundamental theoretical progress and rapid developments in applications across various imaging disciplines. At the same time, a considerable gap remains between the theoretical understanding of the phase retrieval methods and their practical implementation in real-world scenarios. Recognizing this challenge, the recent workshop \textit{Phase Retrieval in Mathematics and Applications (PRiMA)}, held August 5--9 2024 at the Lorentz Center in Leiden, The Netherlands, brought together experts from a diverse set of theoretical and practical backgrounds to explore critical limitations in phase retrieval, and to formulate essential open problems in the field.

As a direct outcome of \textit{PRiMA}, this paper has a twofold purpose: first, to clearly articulate open problems and their practical relevance, thus providing practitioners with insights into theoretical challenges and potential methodological improvements; and, second, to present theorists with concrete, real-world scenarios that need further theoretical investigation and innovation. By bridging this gap, we hope to stimulate productive collaboration and inspire future research directions.

\subsection{Organization of the Paper}
The remainder of this paper is organized as follows. 
Section~\ref{S:Notation} introduces the notation and mathematical framework that will be used in the subsequent sections. 
Section~\ref{S:IntrinsicDifficulty} covers the topic of the \textit{\nameref{S:IntrinsicDifficulty}}, examining complexity-theoretic aspects of the problem.  It surveys measures of problem difficulty, theoretical results on uniqueness and algorithmic hardness, and approaches for establishing confidence in reconstructions.
Section~\ref{S:PhysicsInformedModeling} addresses the problem of \textit{\nameref{S:PhysicsInformedModeling}}, discussing how to incorporate prior knowledge into phase retrieval models. It explores object parameterization strategies (including examples like repetitive unit-cell structures in crystallography and patterned components in imaging) and the challenges of jointly estimating object and probe.  
Section~\ref{S:PtyGenography}, titled \textit{\nameref{S:PtyGenography}}, presents the use of generative models for phase retrieval.  It describes methodologies for leveraging neural-network-based priors, provides examples of reconstructions under varying noise levels, and discusses the observed trade-offs and performance of these methods. 
Section~\ref{S:StructuredRandomPhaseRetrieval} focuses on \textit{\nameref{S:StructuredRandomPhaseRetrieval}}, comparing classical random measurement approaches with structured schemes.  It highlights recent advances in designing practical measurement patterns (for example, block-coded masks or correlated illuminations) that approximate the advantages of randomness. 
Section~\ref{S:OneBitQuantization} investigates \textit{\nameref{S:OneBitQuantization}}, illustrating how extreme data quantization (down to single-bit per measurement) in high-speed electron microscopy impacts phase retrieval and what strategies might mitigate the information loss in such scenarios. 
Finally, Section~\ref{S:MathematicalFoundationsOfWDD} delves into the \textit{\nameref{S:MathematicalFoundationsOfWDD}}, an alternative formulation of phase retrieval, and outlines how analyzing the Wigner distribution (or short-time Fourier transform) can provide insight into unique recovery criteria.

\section{Notation}
\label{S:Notation}

We begin by introducing some general notation that is used in later sections.

\begin{itemize}[leftmargin=*]
\item We denote the imaginary unit by $\ii := \sqrt{-1}$.
\item Euler's number is denoted $e := 2.718\dots$.
\item We write $\mathbb{N}=\{0,1,2,\ldots\}$.
\item For a given positive integer $n$ we denote $[n] := \{ 0, 1, 2, \dots, n-1\}$.
\item Vectors are denoted by bold font small letters, e.g. for ${\bf x} \in \mathbbm{C}^n$. Its $j^{\rm th}$ entry is denoted $x_j \in \mathbbm{C}$, that is, ${\bf x} = \left(x_j \right)_{j\in [n]}$. Note that indexing of vector coordinates starts with $0$.
\item For $1\leq p\leq \infty$ we use $\|{\bf x}\|_p$ to denote the $\ell^p$-norm of ${\bf x} \in \mathbbm{C}^n$.
\item $\mathbb{S}^{n-1} = \{{\bf x}\in \mathbb{C}^n \colon \Vert x \Vert_2 = 1\}$ denotes the unit sphere in~$\mathbb{C}^n$.
\item For two vectors ${\bf x}, {\bf y} \in \mathbbm{C}^n$, their inner product is defined as $\langle {\bf x}, {\bf y}\rangle = \sum_{j\in [n]}x_j\overline{y_j}$.
\item For two vectors ${\bf x}, {\bf y}\in \mathbbm{C}^n$, their coordinate-wise (or Hadamard) product is defined as ${\bf x} \circ{\bf y} = \left(x_j y_j\right)_{j\in [n]}\in \mathbbm{C}^n$.
\item Matrices are denoted by capital letters, e.g., $A \in \mathbbm{C}^{m \times n}$. Its $(j,k)^{\rm th}$ entry is denoted by $A_{jk} \in \mathbbm{C}$, so that $A = \left(A_{jk} \right)_{j\in [m], k\in [n]}$.
\item For a matrix $A \in \mathbbm{C}^{m \times n}$, its adjoint matrix $A^*$ is defined as $A^*_{jk} = \overline{A_{kj}}$.
\item $\text{N}(\mu,\sigma^2)$ denotes the normal distribution with mean $\mu$ and variance $\sigma^2$, $\text{Unif}[a,b]$ the uniform distribution on the interval $[a,b]$, and $\text{Pois}(\lambda)$ the Poisson distribution with intensity $\lambda>0$. 
\end{itemize}

While some of the problems consider an infinite-dimensional setup, where the object and the probe are functions, other problems deal with a discretized, finite-dimensional setup that assumes both of them to be vectors. Here, we introduce the notation for both of these setups.

\subsubsection*{Finite-dimensional setup}
\begin{itemize}[leftmargin=*]
\item The input dimension is denoted by $n\in\mathbb{N} \setminus \{0\}$.
\item The number of measurements is denoted by $m\in\mathbb{N} \setminus \{0\}$.
\item The object to be recovered in a phase retrieval problem is denoted by ${\bf f} \in \mathbbm{C}^n$.
\item A probe (also called mask or window) is denoted by ${\bf w} \in \mathbbm{C}^n$.
\item The $n \times n$ unitary Discrete Fourier Transform matrix is denoted by $F \in \mathbbm{C}^{n \times n}$. Its entries are given by $F_{k,j} := e^{- 2 \pi \ii k j / n}/\sqrt{n}$, for $k, j \in [n]$.
\item For two vectors ${\bf x}, {\bf y}\in \mathbbm{C}^n$, their convolution is defined as $ ({\bf x} \ast {\bf y})_j  = \sum_{k\in [n]}x_ky_{j-k}$, where $j - k$ is understood modulo $n$.
\item For a given ${\bf x}\in\mathbbm{C}^n$, $D_{{\bf x}}\in \mathbbm{C}^{n\times n}$ denotes the diagonal matrix with ${\bf x}$ on its diagonal. In particular, $D_{{\bf x}}{\bf y}={\bf x} \circ {\bf y}$ for ${\bf y}\in\mathbbm{C}^n$. 
\item The shift operator $S_r$ is defined by $(S_r {\bf x})_j = x_{j - r}$, for $j\in [n]$.
\item The modulation operator $M_k$ is defined by $(M_k {\bf x})_j = e^{2 \pi \ii jk / n}x_{j}$, for $j\in [n]$.
\item The measurement matrix is usually denoted by $A\in \mathbbm{C}^{m \times n}$ and the corresponding phaseless measurement map is $\mathcal{A}({\bf f}) = \vert A{\bf f}\vert^2$, where the absolute value is taken coordinate-wise. 
\end{itemize}

\subsubsection*{Infinite-dimensional setup}
\begin{itemize}[leftmargin=*]
\item The object to be recovered is denoted by $f:  \mathbbm{R}^d \rightarrow \mathbbm{C}$, where $d$ is the domain dimension. In most of the cases considered in this paper, we have $d = 2$, but it can also be $d = 1$ or $d = 3$ in some cases.
\item A probe (also called a mask or a window) is denoted by $w:  \mathbbm{R}^d \rightarrow \mathbbm{C}$, where $d$ is the dimension.
\item The Fourier transform of a function $f$ is denoted by $\widehat{f}(\boldsymbol{\xi}) = \mathcal{F}[f] (\boldsymbol{\xi}) = \int_{\mathbbm{R}^d} f({\bf x}) e^{- \ii \langle {\bf x} , \boldsymbol{\xi}\rangle} ~d{\bf x}$ for $d \geq 1$.
\item For two functions $f, g \colon  \mathbbm{R}^d \rightarrow \mathbbm{C}$, their convolution is defined as $(f \ast g)({\bf t})  = \int_{\mathbbm{R^d}} f({\bf \tau})g({\bf t} - {\bf \tau}) ~d {\bf \tau}$.
\item The shift operator $S_{\bf k}$ is defined by $S_{\bf k} f({\bf x}) = f({\bf x} - {\bf k})$.
\end{itemize}

We will now begin to present the individual problems proposed and discussed by various working groups of physicists and mathematicians at PRiMA 2024.

\section[Intrinsic Difficulty of Phase Retrieval]{Problem: Intrinsic Difficulty of Phase Retrieval}
\label{S:IntrinsicDifficulty}

\emph{Authors: Marc Allain, Selin Aslan, Wim Coene, Julien Flamant, Mark Iwen, Oleh Melnyk, Andreas Menzel, Viktor Nikitin, Gerlind Plonka.}
\vspace{0.3cm}

In this section, we consider the general continuous phase retrieval problem
\begin{equation}\label{eq: intensity measurements continuous}
I = \mathcal{A}(f) = \vert A f\vert^2,
\end{equation}
of recovering $f$ from $I$ and its discretization
\begin{equation}\label{eq: intensity measurements discrete}
{\bf I} = \mathcal{A}({\bf f}) = \vert A{\bf f}\vert^2.
\end{equation}
Note that we use the same notation for measurement operator $\mathcal A$ and linear transform $A$ in continuous and discrete cases as they can be distinguished from notation ${\bf f}$ or $f$. We focus mainly on the standard coherent diffraction imaging (CDI) with $A = F$ and ptychography, where $A$ corresponds to the (subsampled) short-time Fourier transform matrix.

In practice, the expected behavior of phase retrieval algorithms depends largely on the sample properties such as support and (in)homogeneity of the phase of the object.
Numerical and experimental evidence shows that all samples are not equally difficult to retrieve. For instance, it is well-known that objects with smooth edges or strong inhomogeneous phase-fields in their domain are generally harder to reconstruct \cite{Fienup1987,Barnett2022, PhysRevLett.121.256101, Wang_2020}. 
Unfortunately, the specifics of the object are not known prior to the experiment. Thus, establishing complexity measures based solely on the measurements is a crucial issue in phase retrieval. 
Possible solutions to this open problem are now discussed, which are mainly structured by the following questions.

\begin{questions}\label{q: intrinsic difficulty}\mbox{}
\begin{enumerate}
\item What are the current practices and what measures could be used for estimating the object's complexity from the measurements?  
\item Which aspects of object complexity are theoretically understood and what open problem should be addressed?
\item How does an object's complexity constitute itself in the reconstruction process? What is its impact on the convergence rate?  
\item Can we justify the obtained reconstruction by providing (pixel-wise) confidence intervals?
\end{enumerate}
\end{questions}

\subsection{Complexity measures}



Let us consider the intensity of the field diffracted by a sample $f$, i.e., 
$I = |F f|^2$, obtained during a 
standard CDI experiment. For the numerical phase retrieval problem, we define the autocorrelation of the discretized problem by
\[
\boldsymbol{\gamma}
= F^{-1} {\bf I} 
\]
where $F^{-1}$ is the inverse discrete Fourier transform operator. We assume hereafter that $\boldsymbol{\gamma}$ corresponds to the spatial sampling of $\gamma$, the autocorrelation of the sample function $f:\mathbb{R}^D \rightarrow \mathbb{C}$. In other words, we assume that the following relation holds
\[
 \boldsymbol{\gamma} = \{ \gamma(\mathbf{r}) : \mathbf{r}\in \mathcal{G} \},
\]
with $\mathcal{G}$ a rectangular $D$-dimensional spatial grid.
%
Since the autocorrelation function and its moments are used in both sparse CDI \cite{Beinert.2017, Novikov.2021} and ptychography \cite{Rodenburg.1992, Chapman.1996, Lazic.2016, Perlmutter.2020, Melnyk.2023b}, its use for object's complexity estimation seems appropriate.
%
Let $S_f$ be the support of $f$,
\[
S_f = \{ \mathbf{r}\in \mathcal{G} :\ f(\mathbf{r}) \neq 0 \}.
\]
%
Then, the support of $\boldsymbol{\gamma}$ is the set of coordinate pairs generated by the elements of $S_f$, \textit{i.e.} $S_\gamma = S_f - S_f$. Thus, in the absence of noise, $\boldsymbol{\gamma}$ is non-zero only within a bounded region 
%
Since ${\bf I}$ is real, $\boldsymbol{\gamma}$ is centrosymmetric and the number of independent constraints provided by $\bf I$ is therefore $\frac{1}{2}|S_\gamma|$, where $\vert S \vert$ denotes the cardinality of the set $S$. Recalling that the number of unknowns is $|S_f|$, the ratio of the number of constraints to unknowns called the \emph{overdetermination ratio} \cite{Elser.2008} is 
\[
\Omega = \frac{1}{2} \frac{|S_f - S_f|}{|S_f|}.
\]
It can be shown that $\Omega \geq 2^{D-1}$, where $D$ is the number of spatial dimensions of the problem. In two dimensions, the overdetermination ratio is always greater or equal to $2$, the equality being verified when the support is convex and centrosymmetric {\cite{Elser.2008}}. Regardless of the structure of the object itself, it can be conjectured that the intrinsic difficulty of phase retrieval depends on this ratio. 
However, $\Omega$ treats all points ${\bf r} \in S_f$ equally, independently of $|f({\bf r})|$. An analogous issue arises with the rank of a matrix, in case of which one can use the stable rank as an alternative \cite{Rudelson.2007}. We question whether a generalized version of $\Omega$ accounting for $|f({\bf r})|$ could be established.

The distribution of autocorrelation values might also provide insight into the difficulty of the phase problem. 
In particular, the normalized second-order moment (also known as the "radius of gyration"), which we define here in the continuous case {\cite{Cohen.1995}}
\[
B := \frac{\int  ||\mathbf{r}||^2 \vert \gamma(\mathbf{r})\vert^2 \mathrm{d}\mathbf{r}}{\int \vert \gamma(\mathbf{r})\vert^2 \mathrm{d}\mathbf{r}},
\]
could be considered under the following assumption
\medskip


{\centerline{\em%
"When $B$ decreases, the difficulty of the phase retrieval problem increases."}}
 \medskip

\noindent 
It is worth noting that $B$ can be given an equivalent expression based on  the diffracted intensity  
\[
B = \displaystyle\frac{\int  || \nabla I(\mathbf{q})||^2 \mathrm{d}\mathbf{q}}{\int \vert I(\mathbf{q})\vert^2 \mathrm{d}\mathbf{q}},
\]
with $\nabla$ being the usual gradient operator. 
Using $B$ as a measure of difficulty is therefore motivated by the following empirical knowledge, disseminated in the CDI community: ``the more fringes in the intensity pattern, the easier the problem''.
Unlike the overdetermination ratio, this metric depends on the value of the autocorrelation, and not only on its support. As such, it might better capture  the intrinsic difficulty of the problem, although it is also more prone to imperfections 
when the autocorrelation is estimated from data (e.g., the autocorrelation estimate can be severely distorted by the noise and/or by the suppression of the central portion of the intensity pattern $I$ induced by a beam-stop). 

%


\subsection{Theoretical background}

During the past decade, significant progress was made in understanding the uniqueness and stability of reconstruction from phaseless measurements. Among them a major step forward was the establishment of the local stability bounds \cite{Alaifari.2019, Cheng.2021, Grohs.2022}. Namely, given two objects $f$ and $g$, it is possible to track their similarity in terms of their diffraction measurements, 
\begin{equation}\label{eq: stability bounds}
\min_{|\alpha|=1} \| f - \alpha g \|_{\mathbb F} \le C(g) \| \mathcal A(f) - \mathcal A(g) \|_{\mathbb M}.   
\end{equation}
Here, the left-hand side is the distance between $f$ and $g$ accounting for global phase ambiguity and the right-hand side is the distance between the respective measurements. Typically, $\| \cdot \|_{\mathbb F}$ is an $\mathbb L^2$-norm, possibly restricted to some domain $\Omega$ and $\| \cdot \|_{\mathbb M}$ is the Sobolev norm, i.e., the sum of $\mathbb L^2$-norms of the function and its derivative. When we only consider functions on a finite domain $\Omega$, then both norms are the $\mathbb L^2$-norms in the corresponding spaces \cite{Grohs.2021}.

The constant $C(g) > 0$ is called the local stability constant as it depends on $g$. This dependency is linked to the connectivity of the support of $g$ represented by the function's Cheeger constant, see \cite{Grohs.2019} for more details.
Thus, when the support of $g$ consists of many dis- or weekly connected components, one can construct a new function $f$ by assigning a separate global phase to each "isle", which will yield similar measurements \cite{Alaifari.2019}.


In the context of investigating the complexity of the given object, it is interesting whether \eqref{eq: stability bounds} could be helpful. For instance, by setting $g = \frac{1}{|\Omega|}\int_{\Omega}f(x) dx$ with $\Omega$ being the domain of $f$, the both sides become analogies to variances. 
If the above idea is applicable, can we then establish bounds as in \eqref{eq: stability bounds} in different spaces $\mathbb F$ to investigate further properties of $f$?  

Another interesting direction in understanding stability is the transversality property introduced in \cite{Barnett2022}. 
For this, let us look at phase retrieval as the intersection task of finding $f \in \mathbb A \cap \mathbb B$. In CDI, $\mathbb A$ is the set of Fourier magnitudes and $\mathbb B$ the support of the object. In ptychography or other diffraction imaging setups with overdetermined linear mapping $A$, we can instead search for $g \in \mathbb A \cap \mathbb B$, where 
\[
\mathbb A := \{ h:\ |h| = y \} 
\quad \text{and} \quad 
\mathbb B:= \{ h:\ h = Af \text{ for some } f\},
\]
are the set of functions satisfying the magnitude constraints and, respectively, the image of the linear operator $A$.   

{When considering $\mathbb B$ as the support constraint of the object,} one of the claims of \cite{Barnett2022} is that the difficulty of phase retrieval mainly depends on the geometry near the intersection $\mathbb A \cap \mathbb B$. In particular, on the dimension of its near-linearization $T_f \mathbb A \cap T_f \mathbb B$, where $T_f \mathbb A, T_f \mathbb B$ are the tangent spaces at $f$ and since $\mathbb B$ is linear, we can write $T_f \mathbb B = \mathbb B$. 
The intersection is said to be transversal if $\dim (T_f \mathbb A \cap \mathbb B) = 0$, and non-transversal otherwise.

One key of the results in \cite[Theorem 3.5]{Barnett2022} is that phase retrieval has a locally Lipschitz inverse mapping at $f$ if and only if the intersection $T_f \mathbb A \cap \mathbb B$ is transversal. 
At the present time, it is unclear (at least for us) whether the notion of transversality can be useful in practice. The difficulty in using it seems to be the direct dependency on $f$ that is not available to us. 

It would also be beneficial to put transversality into perspective of other uniqueness and stability results known for phase retrieval, such as complementarity property \cite{Bandeira.2014, Alaifari.2017} or (robust) rank null space property \cite{Kabanava.2016}.


\subsection{Complexity and reconstruction}

Turning to the reconstruction algorithms, we observe that even less is understood about the impact of an object's complexity on the convergence speed. Let us consider a gradient descent method
\[
{\bf f}^{t+1} = {\bf f}^t - \mu \nabla \mathcal L({\bf f}^t)
\]
applied to the "square-root" objective
\[
\mathcal L ({\bf f}) = \sum_{j=1}^m \big| | (A {\bf f})_j |^{1/2} - {\bf y}_j^{1/2}  \big|^2,
\]
which, from the perspective of maximum likelihood estimation, is an approximation of the Poisson log-likelihood function \cite{Thibault.2012}. 
With appropriately chosen step size $\mu$ \cite{Xu.2018}, its convergence is characterized by 
\begin{equation}\label{eq: convergence rate}
\nabla \mathcal L({\bf f}^t) \to 0 \text{ as } t \to \infty,
\ \text{and} \ 
\max_{t = 1, \ldots, T} \| \nabla \mathcal L({\bf f}^t)  \| \le \frac{(\mathcal L({\bf f}^0) - \inf_{{\bf f}} \mathcal L({\bf f}))^{1/2}}{\| A \| \sqrt{T}},
\end{equation}
where $T$ is the number of iterations and $\| A \|$ is the spectral norm of the operator $A$.

A few important remarks are the following:
\begin{itemize}
    \item By convergence we understand finding a critical point $\nabla \mathcal L({\bf f}) = 0$, which may not necessarily be the global minimizer. For nonrandom $A$, this is the best-guaranteed performance up to date as the function $\mathcal L$ is nonconvex. On the other hand, for random $A$ the landscape of the (slightly modified) objective is well-understood \cite{Cai.2021}. 

    \item If more than one unknown is being optimized at the time, e.g., probe in blind ptychography \cite{Melnyk.2023}, the exponent $1/2$ in the denominator is replaced by $p/2$ where $p$ is the number of unknown components, highlighting how reconstruction difficulty scaled with the problem's difficulty.

    \item It is important to note that \eqref{eq: convergence rate} is a so-called global bound, which applies for an arbitrary starting point ${\bf f}^0$. Theoretical findings in \cite{Bendory.2018} suggest that for ptychographic reconstruction with the right initialization, it is possible to achieve linear convergence rate, that is to replace $T^{-1/2}$ in \eqref{eq: convergence rate} with $e^{- cT}$ for $c>0$. However, how to obtain a sufficiently good initial guess efficiently remains unclear.  On the contrary, in \cite{Huang.2025} for one-dimensional Fourier phase retrieval, the existence of a linearly convergent algorithm is shown to be impossible\footnote{unless P = NP}.  
\end{itemize}

Prominently, in \eqref{eq: convergence rate} the impact of noise, measurement operator $ A$, initialization ${\bf f}^0$ and the complexity of the ground truth is hidden in the term \mbox{$\mathcal L({\bf f}^0) - \inf_{{\bf f}} \mathcal L({\bf f})$}. 


Due to the empirical evidence on the slower convergence for more complex objects, an important direction of research is to establish a more explicit dependency between the convergence rate \eqref{eq: convergence rate} and, for instance, the object's smoothness class. 

An alternative idea is to perform a numerical study on the conditioning of the Hessian $\nabla^2 \mathcal L$ as it determines the (local) convexity of the function and the convergence rate of the gradient method. This leads to a natural follow-up question whether the desired complexity statistics of the object can be based on the Hessian and its conditioning?    

\subsection{Confidence intervals}

Adjacent to the Hessian is the topic of confidence intervals and uncertainty quantification for reconstructed images or their parts. 
The classical solution to this problem is the construction of confidence intervals, within which the ground truth should lie compared to the obtained estimate. The construction of confidence intervals is particularly crucial for medical applications, where false reconstruction may pose dangers to human health.

To construct the intervals for ${\bf f}$ using results for maximum likelihood estimation we require a Fisher information matrix, which is obtained by taking an expectation of the Hessian matrix. This can be done, following the ideas from \cite{Wei.2020}. However, the performance of the method is limited to dimensions below 100 pixels by the computation time of the inverse of the Fisher information matrix. Whether the computations can be sped up, e.g., using techniques from randomized linear algebra \cite{Martinsson.2020}, is an open question.

The confidence interval construction is not limited to the maximum likelihood method.
Another possibility to compute confidence intervals is using conformal inference techniques \cite{Angelopoulos.2021}. This approach, however, requires a large dataset of possible objects for calibration.  Unfortunately, a large realistic dataset of objects is not yet available in diffraction imaging, which significantly limits the applications of conformal inference.  The assembly of such a dataset is a high-value future research objective.

\section[Physics-Informed Modeling of Object and Probe]{Problem: Physics-Informed Modeling of Object and Probe}
\label{S:PhysicsInformedModeling}

\emph{Authors: Selin Aslan, Oleh Melnyk, Palina Salanevich, Mark Iwen, Felix Krahmer, Gerlind Plonka.}

\vspace{0.3cm}

With the advancement of detector capabilities, the dimensionality and complexity of measurement models in diffraction imaging continue to grow, leading to a significant increase in computational demands. Estimating millions of unknown parameters in such high-dimensional models often creates a bottleneck, where iterative methods can stagnate as the updates for each individual pixel are close to zero. To overcome this limitation, reducing the number of unknowns through suitable parameterization of the underlying objects offers a promising solution.

\begin{figure}[h]
\subfloat[Simulation of SrTiO$_3$ crystal]{
\includegraphics[width = 0.45\textwidth]{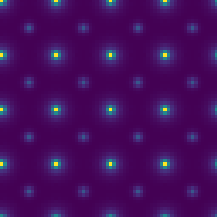}%
}
\subfloat[2D slice of a simulated chip]{
\includegraphics[width = 0.45\textwidth]{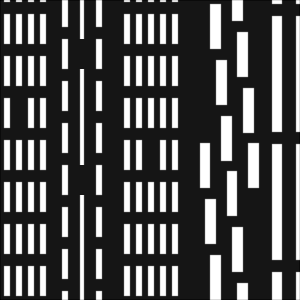}%
}
\caption{Examples of objects. \label{fig: parametrization example}}
\end{figure}

In many applications, the specimen exhibit specific structure that can be represented as a compilation of repeating patterns. 
For example, in crystallography \cite{Chen.2021, Diederichs.2024}, crystals consist of layers with repeating arrangements of individual atoms. 
Similarly, in circuit board imaging \cite{Shao.2024}, the focus is on reconstructing patterns formed by numerous individual components, each belonging to a finite set of admissible types, such as chips, resistors, capacitors, and connections. Visualizations of such objects can be found in \Cref{fig: parametrization example}.

The idea of parametrization is not necessarily restricted to the object itself and can be used for other experimental unknowns. For instance, in electron microscopy, the probe is typically represented by aberrations of circular aperture in terms of Zernike polynomial expansion \cite{Mahajan.1994, Ou.2014}.

Note that the idea of efficient parametrization is not new to imaging and phase retrieval. Ideas from compressed sensing \cite{Jaganathan.2013, Beinert.2017, Iwen.2017, Wang.2018}, dictionary learning \cite{Tillmann.2016, Krishnan.2018}, and, generative learning \cite{Aslan.2021, Fahad.2021, Seifert.2024} have been explored in the past. 
Among these, generative models had the most success in phase retrieval so far, both with pre-trained  \cite{Aslan.2021, Seifert.2024} and zero-shot \cite{Fahad.2021} networks. However, the physical interpretability of such models is limited, which motivates us to consider model-based parametrizations.

\subsection{Object Parameterization}
Motivated by the applications in crystallography and lithography, let us assume that the object $f$ can be modeled as a sum of complex-valued measures
\begin{equation}\label{eq: distribution model general}
f = a_0 \lambda + \sum_{j=1}^N a_j S_{t_j} \mu_{j}(\sigma),
\end{equation}
\noindent where $\lambda$ 
denotes the Lebesgue measure representing the 
general object background (i.e., absence of features), and the measures $\mu_{j}$ are (normalized, localized) complex measures representing the features of the object. We assume that $\mu_j \in \mathcal{E}$, where $\mathcal E$ with $|\mathcal E | \ll N$ is a set of all admissible features. Moreover, the measures $\mu_j$ may be influenced by experimental conditions captured by the parameter $\sigma$.
In this model, the translation operator $S_{t_j}$ centers $\mu_j$ at the position~$t_j$. The weights $a_j$ represent the strength of the $j$th feature. 

In the applications above, measures $\mu_j$ represent atomic potentials in crystallography and individual chip components in circuit boards. For example, let us consider the crystallographic model from \cite{Diederichs.2024}, 
where $f$ represents the Coulomb potential, incorporating temperature effects through the Debye-Waller damping scheme. 
The model is expressed as
\begin{equation}\label{eq: signal model}
f = \exp \left(\ii\sigma\sum_{j = 1}^N  \omega_j S_{t_j} v_j\right),
\end{equation}
\noindent where $\sigma$ is the interaction constant, $v_j$ is the potential function of the $j$th atom, $\omega_j$ is the weight and $t_j$ is the atom's position. If the atoms are sufficiently far apart and the supports of their potentials are approximately disjoint,
we can set $\omega_j = 1$ and rewrite \eqref{eq: signal model} in the form of \eqref{eq: distribution model general},
\[
f \approx \lambda + \sum_{j = 1}^N S_{t_j}(1 -  \exp (\ii\sigma v_j))
\approx
\lambda + \sum_{j = 1}^N S_{t_j}(1 -  \exp (\ii\sigma  v_j(0))) .
\]

A crucial detail is that $\mathcal E$ is known, as the individual potentials $v \in \mathcal E$ of atoms are computed in \cite{Lobato.2014}. 
In this particular application, the authors initially perform a pixel-based reconstruction, from which they determine $N$, $\mu_j$ and set up initial guesses for $\sigma$, $a_j$, and $t_j$, which are then optimized.

The above example raises a few important questions.

\begin{question}\label{q: dictionary selecion}
The set $\mathcal E$ strongly depends on the application and may not be known in advance. What is an efficient strategy to establish it?
\end{question}
Here, we see a potential solution in using patch-based priors \cite{Tillmann.2016, Altekruger.2023} to learn $\mathcal E$. However, this approach is prone to reliability and interpretability problems with the learned dictionary.

\begin{question}\label{q: parameter selecion}
How can one select the number of features $N$ and the features~$\mu_j$?
\end{question}
As demonstrated in~\cite{Diederichs.2024}, one approach to this problem is to use the pixel-wise reconstruction as a way to transit to a parametrized model. 
It can be further developed by considering of a coupled model, where the parametrization model is used as a prior or a regularization term akin to \cite{Tillmann.2016}. Studying this idea, as well as alternative methods for selecting $N$ and $\mu_j$, is an important component in development of an efficient phaseless reconstruction method.

\begin{question}\label{q: unique recovery}
Can we establish uniqueness and stability guarantees for the model~\eqref{eq: distribution model}? What are the reasonable sufficient conditions?
\end{question}

In the model \eqref{eq: distribution model general}, the set of parameters is given by  $(a_j, \mu_j, t_j)$, $1 \le j \le N$ and is generally much smaller than in the pixel-wise representation of the object. While this approach effectively reduces the number of parameters, it comes at the cost of nonlinear interactions between $a_j$, $\mu_j$ and $t_j$, which makes the analysis of the model more intricate.
We discuss Question~\ref{q: unique recovery} in more detail for a simplified model in Section~\ref{sec: sparse crystallography}.

\subsection{Sparse Crystallography}\label{sec: sparse crystallography}
In this section, we focus on the following  simplification of the crystallographic model \eqref{eq: signal model}. We replace the atomic potentials $v_j$ with the Dirac delta measure $\delta$. In this way, we disregard the weight of an atom and rather focus on its position. Then,~\eqref{eq: signal model} can be rewritten in terms of distributions as 
\begin{equation}\label{eq: distribution model}
f = 
\lambda + \sum_{j = 1}^N \left(e^{\ii \omega_j} - 1\right) \delta_{t_j} = \lambda + \sum_{j=1}^N a_j \delta_{t_j},
\end{equation}
\noindent where $a_j = e^{\ii \omega_j} - 1$ for $j\in \{1,\dots, N\}$.

\begin{problem}\label{pr: sparse crystallography}
Suppose the signal $f$ is given by the model~\eqref{eq: distribution model} and let $w\in\mathcal{S}(\mathbbm{R}^2)$ be the probe. Given the measurements of the form 
\begin{equation}\label{eq: measurements}
\mathcal{A}(f)(r, \xi) = \vert \mathcal{F}(f\cdot S_r w)(\xi)\vert^2, \quad \xi\in \Omega,~r\in R,
\end{equation}
\noindent recover the atom positions $t_j$ and weights $a_j$ for $j \in \{1, \dots, N\}$.
\end{problem}

\begin{remark}
In the case of no background, that is, when $f = \sum_{j=1}^N a_j \delta_{t_j}$, the problem can be seen as typical sparse phase retrieval, which can be solved using the Prony's method \cite{Beinert.2023} or lifting with TV-regularization \cite{Abouserie.2022}.
\end{remark}

Note that by replacing the atomic potentials with the Dirac delta measure, we transform the object model into a discrete representation. To further simplify the problem, let us now discretize it and consider the finite-dimensional version of ~\eqref{eq: distribution model}

\begin{equation}\label{eq: discrete model}
{\bf f} = 
{\bf\mathbbm{1}} + \sum_{j = 1}^N a_j {\bf e_{t_j}}\in \mathbb{C}^n.
\end{equation}
Then the measurements~\eqref{eq: measurements} become

\begin{align}
\mathcal{A}(f)(r, k) 
& = \left\vert 
(FS_r {\bf w})_k + \sum_{j = 1}^N a_j(S_r{\bf w})_{t_j}(F{\bf e_{t_j}})_k\right\vert^2 \nonumber \\ 
& = \left\vert 
(FS_r {\bf w})_k + \sum_{j = 1}^N a_j(S_r{\bf w})_{t_j}F_{k,t_j}\right\vert^2 \nonumber \\ 
& = \left\vert 
(M_{-r}F {\bf w})_k + \sum_{j = 1}^N a_j(S_r{\bf w})_{t_j}F_{k,t_j}\right\vert^2 \nonumber \\ 
& = \left\vert 
(F {\bf w})_k + \sum_{j = 1}^N \frac{a_j}{\sqrt{n}}(S_r{\bf w})_{t_j}e^{-2\pi \ii k(t_j - r)/n}\right\vert^2. \label{eq: affine pr}
\end{align}
Note that the background term $
(F {\bf w})_k$ here does not depend on the probe shift~$r$. As a result, the above problem could be treated as a special case of affine phase retrieval \cite{Gao.2018}.  

This consideration leads to the following open questions, answering which would allow us to understand benefits and limitations of the proposed model-based object parametrization.

\begin{question}\label{q: number of shifts}
How large the set of the probe shifts $R$ should be to ensure unique reconstruction of $a_j$, $t_j$, $j\in \{1,\dots, N\}$ from~\eqref{eq: affine pr}?
\end{question}

\begin{question}\label{q: methods} What methods can be used to solve this problem efficiently and stably, and under what conditions on the probe ${\bf w}$ and the set of the probe shifts $R$?
\end{question}
To further improve the reconstruction accuracy and/or to reduce the cardinality of $R$, one can incorporate prior knowledge on $a_j$, $t_j$. The examples of such priors are the admissible range of $\vert a_j\vert$ and the separation $\vert t_j - t_\ell\vert \geq \delta$ for $j\neq \ell$ and known $\delta > 0$. Furthermore, in crystallographic applications, the atom positions $t_j$ often form a superposition of several latices.

\section[PtyGenography]{PtyGenography: Using Generative Model for Regularization with Varying Noise Levels}
\label{S:PtyGenography}

\emph{Authors: Selin Aslan, Wim Coene, Sjoerd Dirksen, Tristan van Leeuwen, Allard Mosk, Palina Salanevich.}
\vspace{0.3cm}

In this section, we consider a more general setup that includes the phase retrieval problem as a partial case.
Namely, we consider a non-linear inverse problem of retrieving ${\bf f}\in \mathbb{C}^n$ from noisy measurements ${\bf y} = \mathcal{A}({\bf f}) + \boldsymbol{\varepsilon}$, where $\mathcal{A}\colon \mathbb{C}^n \to \mathbb{C}^m$ is the measurement map, and $\boldsymbol{\varepsilon}\in \mathbb{C}^m$ is the noise term. 
In the case of the phase retrieval problem, the measurement map is defined as $\mathcal{A}({\bf f}) = \vert A {\bf f} \vert^2$, where $A\in\mathbb{C}^{m\times n}$ is a given matrix (see, e.g. \cite{Grohs2020, Dong2023} for a recent overview). Note that in this section, we only consider the additive noise model, which is typical for detector noise and does not cover Poisson shot noise.

Assuming $\mathcal{A}$ is injective, the conventional formulation of the reconstruction problem is
\begin{equation}\label{eq: conventional}
 \min_{{\bf f}} \|\mathcal{A}({\bf f})-{\bf y}\|_2^2.
\end{equation}
To account for instabilities in the presence of noise, one often adds a regularization term. The most common, classical regularization approach is Tikhonov regularization, which is extensively studied and theoretically well-understood~\cite{vasin2006some}. 
However, it tends to smooth out high-frequency components, which can be problematic when trying to capture detailed features of the signal.

More recently, generative models have emerged as a powerful alternative to capture detailed prior information on the signal, and, thus, enforce reconstruction stability. More specifically, assuming that the signal ${\bf f} = \mathcal{G} ({\bf z})$ for some ${\bf z}\in \mathbb{C}^k$, where $\mathcal{G}\colon \mathbb{C}^k \to \mathbb{C}^n$ is a given generative model and $k<n$, the problem \eqref{eq: conventional} becomes
\begin{equation}\label{eq: generative}
\min_{\bf z} \|\mathcal{A}\circ \mathcal{G}({\bf z})-{\bf y}\|_2^2.
\end{equation}
The rationale here is that the conditioning of $\mathcal{A}\circ\mathcal{G}$ is generally more favorable than that of $\mathcal{A}$, at the expense of introducing a bias in the reconstruction. It has indeed been observed in numerical experiments that for high signal-to-noise ratio levels formulation~\eqref{eq: conventional} performs better, while for in the case of low signal-to-noise ratio formulation~\eqref{eq: generative} outperforms it \cite{Seifert2024}. 

\medskip

This section provides an overview of the recent work by the authors~\cite{Aslan2025} on regularizing an inverse problem using generative priors and discusses the open problems that emerge from it. This work~\cite{Aslan2025} was initiated and partially conducted during the  PRiMA workshop at the Lorentz center.

In this section we
\begin{enumerate}
    \item characterize the reconstruction error of \eqref{eq: conventional} and \eqref{eq: generative} in terms of the bias and variance;
    \item develop a practical method for detecting bias introduced by generative priors;
    \item propose a unified variational framework that balances generative and conventional formulations, achieving stable performance across different noise levels.
\end{enumerate}

In Section~\ref{method}, we explore the recovery properties of the two inverse problem formulations~\eqref{eq: conventional}~and~\eqref{eq: generative}. Section~\ref{sec: reconstruction error} presents the theoretical foundations obtained in~\cite{Aslan2025}, including the key results on the reconstruction error 
and the bias introduced by the generative model that may hinder recovery in the cases when the true signal lies outside the learned distribution. A unified approach that allows to mitigate this problem and balance generalization and stability by combining~\eqref{eq: conventional}~and~\eqref{eq: generative} is discussed in Section~\ref{sec: unified approach}. Section~\ref{results} summarizes numerical findings that illustrate the trade-offs between generative and classical approaches. Finally, Section~\ref{conclusion} discusses open problems and future research directions, emphasizing challenges in bias detection, theoretical guarantees, and optimization strategies.

\subsection{Methods}\label{method}
We assume that both the measurement map \(\mathcal{A}\) and the generative model \(\mathcal{G}\) are injective and bi-Lipschitz with constants \(\alpha, \beta \geq 1\). That is, for any \({\bf f}, {\bf f'} \in \mathbb{C}^n\),
\[
\alpha^{-1}\|{\bf f}-{\bf f'}\|_2 \leq \|\mathcal{A}({\bf f}) - \mathcal{A}({\bf f'})\|_2 \leq \alpha \|{\bf f}-{\bf f'}\|_2,
\]
and likewise for $\mathcal{G}$, with $\beta$ in place of $\alpha$.

\begin{remark}
For the phase retrieval problem, where \(\mathcal{A}({\bf f}) = |A {\bf f} |^2\) with \(A \in \mathbb{C}^{m \times n}\), the measurement map \(\mathcal{A}\) satisfies the (bi-)Lipschitz property whenever it is injective \cite{balan2016lipschitz}. However, for most matrices \(A\), no explicit bound on the Lipschitz constant \(\alpha\) is known. An exception is phase retrieval from locally supported measurements \cite{iwen2019lower}.
\end{remark}

We furthermore assume that the bi-Lipschitz constant $\gamma$ of $\mathcal{A}\circ\mathcal{G}$ is more favorable than that of $\mathcal{A}$, that is, $\gamma < \alpha$. In other words, $\mathcal{G}$ effectively regularities the inverse problem. Moreover, we assume that $\mathcal{G}$ is well-conditioned, in the sense that $0 < \beta - 1 \ll 1$.

\subsubsection{Characterizing the reconstruction error}\label{sec: reconstruction error}
We denote by ${\bf f_0}$ the ground-truth signal we aim to recover from the measurements
\[
{\bf y} = \mathcal{A}({\bf f_0}) + \boldsymbol{\varepsilon},
\]
where $\boldsymbol{\varepsilon}$ is a bounded noise term. The following results characterize the reconstruction error for conventional~\eqref{eq: conventional} and generative priors-based~\eqref{eq: generative} reconstruction methods~\cite{Aslan2025}.

\begin{lemma}
Let $\displaystyle{\bf \tilde{f}} = \arg\min_{{\bf f}} \|\mathcal{A}({\bf f})-{\bf y}\|_2$ with ${\bf y} = \mathcal{A}({\bf f_0}) + \boldsymbol{\varepsilon}$. Then the reconstruction error is given by 
\[
\|{\bf \tilde{f}} - {\bf f_0}\|_2 \leq 2\alpha \|\boldsymbol{\varepsilon}\|_2.
\]
\end{lemma}

\begin{lemma}\label{lem: generative bound}
Let $\displaystyle{\bf \tilde{z}} = \arg\min_{\bf z} \|\mathcal{A}\circ\mathcal{G}({\bf z}) - {\bf y}\|_2$ with ${\bf y} = \mathcal{A}({\bf f_0})+\boldsymbol{\varepsilon}$, and ${\bf\tilde{f}}=\mathcal{G}({\bf\tilde{z}})$. Then the reconstruction error is bounded by 
\[
\|{\bf \tilde{f}} - {\bf f_0}\|_2 \leq (1 + 2\alpha\beta\gamma) \|\mathcal{G}({\bf z_0}) - {\bf f_0}\|_2 + 2\beta\gamma\|\boldsymbol{\varepsilon}\|_2,
\]
where $\displaystyle{\bf z_0} = \arg\min_{\bf z} \|\mathcal{G}({\bf z})- {\bf f_0}\|_2$.
\end{lemma}

Note that under the assumptions stated above, it is not unreasonable to have $\beta\gamma < \alpha$, so that the regularized problem indeed leads to less amplification of noise, at the expense of introducing a bias which mainly depends on the expressiveness of the generative model. 
More specifically, there is a trade-off between the introduced bias and the noise amplification. More restrictive generative model would have smaller $\beta$, leading to the smaller variance, but at the same time it would lead to large bias term in out-of-distribution scenarios. This phenomenon is illustrated in Figure~\ref{fig:example1_recon}.

\subsubsection{Detecting bias}\label{sec: bias}
A significant challenge in phase retrieval is detecting and correcting bias due to imperfect measurements. 
In many applications, such as material science~\cite{agour2013investigation, lehmann2022characterization}, lithography~\cite{mochi2010actinic}, and circuits board manufacturing~\cite{kaya2023development}, the ground truth signal has the form ${\bf f_0} = \mathcal{G}({\bf z_0}) + {\boldsymbol{\eta}}$, where $\mathcal{G}$ models all the ``perfect'' signals and ${\boldsymbol{\eta}}$ represents the signal imperfections and manufacturing defects.
Detecting these defects is critical for the success of the reconstruction process. Therefore, we would like to answer the following question.

\begin{question} Given a solution $\displaystyle{\bf \tilde{z}} = \arg\min_{\bf z} |\mathcal{A}\circ\mathcal{G}({\bf z}) - {\bf y}|_2$ to the optimization problem~\eqref{eq: generative}, can we detect or perhaps even quantify the bias $|\mathcal{G}({\bf \tilde{z}})- {\bf f_0}|$? \end{question}

Answering this question requires more refined exploration of the properties of the measurement map $\mathcal{A}$. One possible approach to it is using \emph{Vanderlugt correlation}~\cite{Lugt1964}, which gives us a way to highlight the defect structure by correlating $\mathcal{A}\circ\mathcal{G}({\bf \tilde{z}})$ with ${\bf y}$.

\subsubsection{Vanderlugt correlation}\label{sec: Vanderlugt correlation}

In this section, let us assume that the measurements are diffraction patterns in the Fraunhofer regime, so that the measurement map is given by $\mathcal{A}({\bf f}) = \vert F{\bf f}\vert^2$, where $F$ is a discrete Fourier matrix~\cite{Goodman1988}. In this setup, a procedure inspired by the \emph{VanderLugt filter} provides a simple means to detect and highlight object defects. 
The analog VanderLugt filter\cite{Lugt1964} can detect a given object by correlation in the Fourier domain, using an analog or digital hologram of the object to be recognized. Here, we adapt this principle to find and highlight the imperfections ${\boldsymbol{\eta}}$, assuming these are spatially sparse.

We consider the Fourier transform \( F{\bf f} \) of the optimized ``perfect'' object ${{\bf f}=\mathcal{G}({\bf z_0})}$ obtained by fitting the ground truth ${\bf f_0}$ to the generative model. For the diffraction pattern intensity, we then have
\[
|F{\bf f_0}|^2 = |F{\bf f}|^2 + |F{\boldsymbol{\eta}}|^2 + F{\boldsymbol{\eta}}(F{\bf f})^* +(F{\boldsymbol{\eta}})^* F{\bf f}.
\]
In the absence of experimental noise, when  ${\bf y} = |F{\bf f_0}|^2$, the solution obtained by the reconstruction method~\eqref{eq: generative} is ${\bf f}$. The term $|F{\boldsymbol{\eta}}|^2$ is quadratic in the object imperfection  ${\boldsymbol{\eta}}$ that we assume to be small, so that this term can be neglected. Then we have
\[
{\bf y} -|F({\bf f})|^2  \approx F{\boldsymbol{\eta}} (F{\bf f})^* +( F \boldsymbol{\eta})^* F {\bf f},
\]
which we can rewrite as 
\[
F{\boldsymbol{\eta}} \approx \frac{{\bf y} -|F{\bf f}|^2 }{ (F{\bf f})^*} - \frac{ (F{\boldsymbol{\eta}})^* F{\bf f}}{ (F{\bf f})^*}.
\]

\noindent The first term on the RHS of the equation can be calculated directly, and is a hologram of the misfit between the ground truth measurements and the measurements of the object recovered by~\eqref{eq: generative}. The second term  is a conjugate hologram. Such terms are common in holography and typically give rise to speckle-like background noise. Using this equation, we can approximate the object imperfection $\boldsymbol{\eta}$ by taking the inverse Fourier transform.

\begin{figure}
\centering
\begin{tabular}{cc}
\includegraphics[scale=.3]{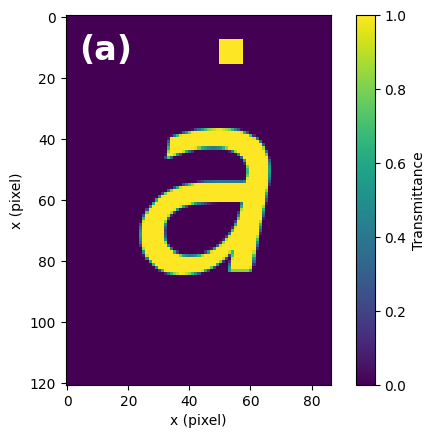}
&
\includegraphics[scale=.3]{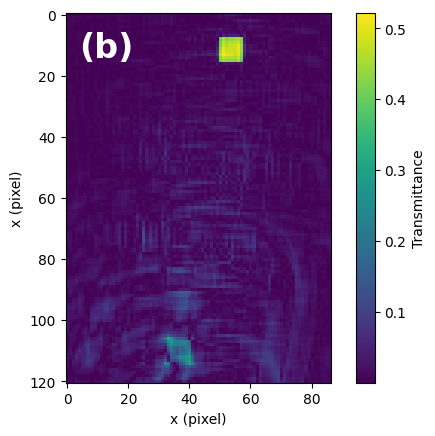}

\end{tabular}
\caption{ (a) Example of an object ${\bf f_0} = {\bf f} + \boldsymbol{\eta}$, where ${\bf f}$ is the letter 'a' and the imperfection $\boldsymbol{\eta}$ is the dot accent. (b) Imperfection ${\boldsymbol{\eta}}$ retrieved from correlating the generative reconstruction ${\bf f}$ with the given phaseless measurements.}
\label{fig:dotted_a}
\end{figure}

To illustrate this consideration numerically, let us assume that the generative model approach~\eqref{eq: generative} perfectly reconstructs letters from the standard roman alphabet, but does not allow accents. Let the ground truth image ${\bf f_0}$ be as in  Fig.~\ref{fig:dotted_a}(a). Since the generative model only reconstructs the letter `a' and not the dot, the procedure outlined above outputs exactly the dot as shown in Fig.~\ref{fig:dotted_a}(b). The conjugate term gives rise to the noisy fringes in the image.

\subsubsection{Towards a unified approach}\label{sec: unified approach}
We propose a combined approach to inference using a generative model
\begin{equation}\label{eq:combined}
\min_{{\bf z},{\bf f}} \|\mathcal{A}({\bf f})-{\bf y}\|_2^2 + \lambda^2 \|\mathcal{G}({\bf z})-{\bf f}\|_2^2.
\end{equation}
Obviously, as $\lambda = 0$, we retrieve \eqref{eq: conventional}, and as $\lambda\rightarrow\infty$, we retrieve \eqref{eq: generative}. In~\cite{Aslan2025}, we show the following reconstruction guarantees for~\eqref{eq:combined}.

\begin{lemma}\label{lem: unified bound}
The estimate ${\bf \tilde f}$ resulting from \eqref{eq:combined} satisfies
\[
\|{\bf \tilde f}-{\bf f_0}\|_2 \leq \lambda\alpha \|G({\bf z_0})-{\bf f_0}\|_2 +  2\alpha\|\boldsymbol{\varepsilon}\|_2
\]
\end{lemma}

\bigskip

We note that can view \eqref{eq: conventional}, \eqref{eq: generative}, and \eqref{eq:combined} as partial cases of the generic non-linear optimization problem of the form
\begin{equation}\label{eq: unified formulation}
\min_{{\bf x}\in \mathbb{C}^d} \|\mathcal{A}\circ \mathcal{B}({\bf x}) - {\bf y}\|_2^2 + \lambda^2 \|{\bf w}\odot {\bf x}\|_2^2.
\end{equation}
We retrieve the specific instances as follows:
\begin{itemize}
\item \eqref{eq: conventional} by letting $d = n$, ${\bf x}\coloneq {\bf f}$, $\mathcal{B} = \mathcal{I}$, and $\lambda = 0$;
\item \eqref{eq: generative} by letting $d = k$, ${\bf x}\coloneq {\bf z}$, $\mathcal{B} = \mathcal{G}$, and $\lambda = 0$;
\item \eqref{eq:combined} by letting $d = k + n$ with ${\bf x} = ({\bf x_1}, {\bf x_2})$, where ${\bf x_1}\in \mathbb{C}^k$ and ${\bf x_2}\in \mathbb{C}^n$. We define $\mathcal{B}:\mathbb{C}^{n+k} \rightarrow \mathbb{C}^n$ as $\mathcal{B}({\bf x_1}, {\bf x_2}) = \mathcal{G}({\bf x_1}) + {\bf x_2}$, and $w\in \mathbb{C}^{k + n}$ is defined by $w(t) = 0$ for $t\in [k]$ and $w(t) = 1$ for $t\in [k + n]\setminus [k]$. The optimization problem~\eqref{eq:combined} then follows by taking ${\bf x_1} \coloneq {\bf z}$ and ${\bf x_2} \coloneq {\bf f} - \mathcal{G}({\bf z})$.
\end{itemize}

To obtain the object reconstruction ${{\bf\tilde f} = \mathcal{B}({\bf\tilde z})}$, we solve~\eqref{eq: unified formulation} using a Quasi-Newton method, such as limited-memory Broyden-Fletcher-Goldfarb-Shanno (L-BFGS) algorithm~\cite{liu1989limited}.
The reconstruction algorithm we employ thus has parameters $\mathcal{B}, \lambda, {\bf w}$, and a stopping tolerance for L-BFGS.

\subsection{Numerical results}\label{results}
For our numerical experiment, we define the measurement map as
\[
\mathcal{A}({\bf f}) = |A{\bf f}|^2,
\]
where matrix $A$ represents a masked Fourier transform, corresponding to the measurements with $\ell$ different probes 
\[
A = \left(\begin{matrix} F\text{diag}({\bf a}_1) \\ F\text{diag}({\bf a}_2) \\ \vdots \\ F\text{diag}({\bf a}_\ell)\end{matrix}\right).
\]
Here, ${\bf a}_i \in \mathbb{R}^{n}$ are random binary probes, that is, their entries are independent identically distributed Bernoulli random variables. The number of measurements is thus $m=n\cdot\ell$.

We define the generative model as
\[
\mathcal{G}({\bf z}) = G{\bf z} + {\bf b},
\]
where ${\bf b}\in\mathbb{C}^n$ and $G\in\mathbb{R}^{n\times k}$
are obtained by principle component analysis of a data set of handwritten digits \cite{deng2012mnist}. The elements of the data set are $8\times 8$ images, so that $n=64$. For our experiment, we choose $k = 30$.

\begin{figure}
\centering
\includegraphics[scale=.6]{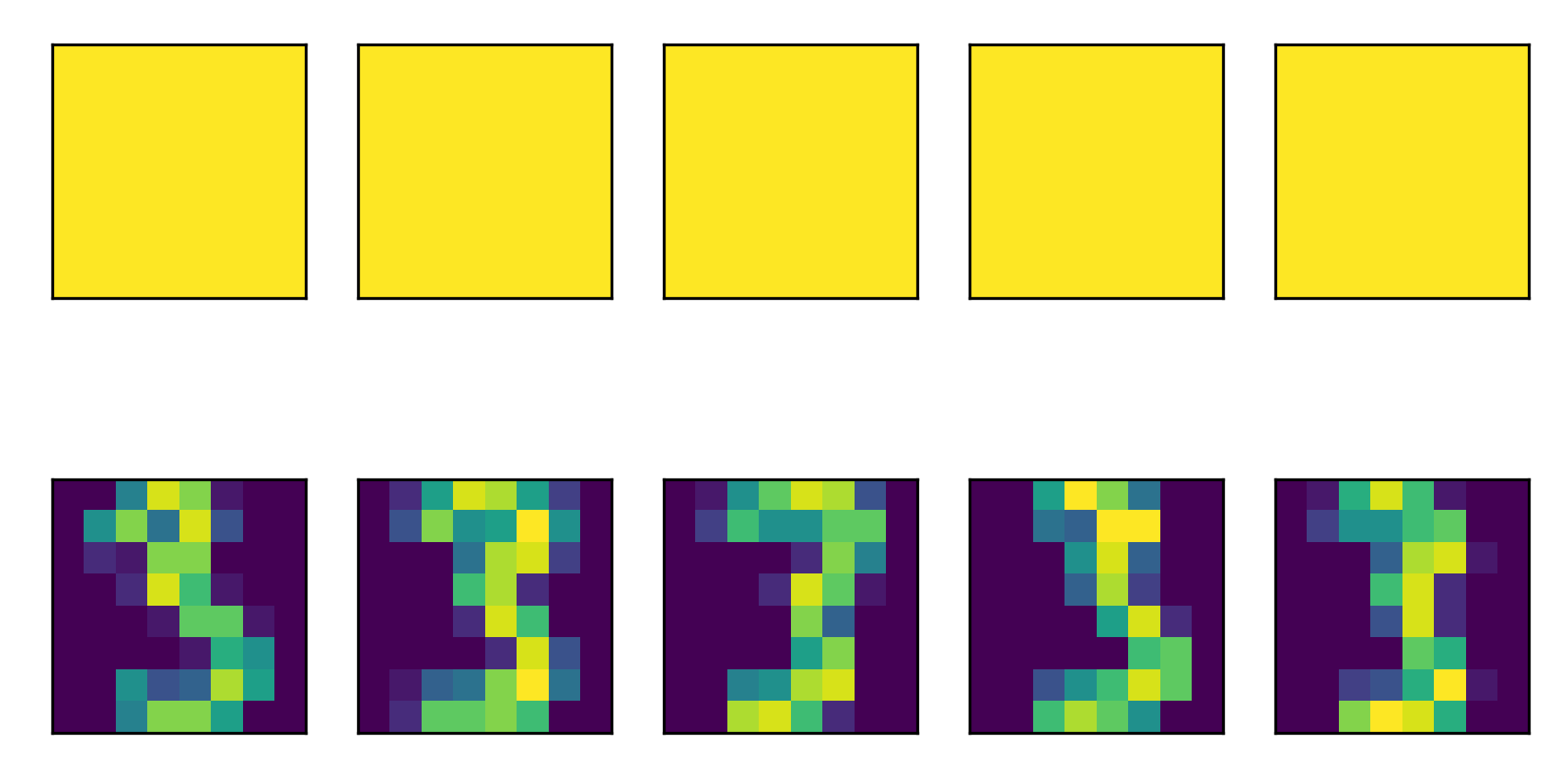}
\caption{Samples of the data set on which the generative model was trained ($n=64$). The top row displays the real part while the bottom row displays the imaginary part.}
\label{fig:example1_data}
\end{figure}
\begin{figure}
\centering
\includegraphics[scale=.6]{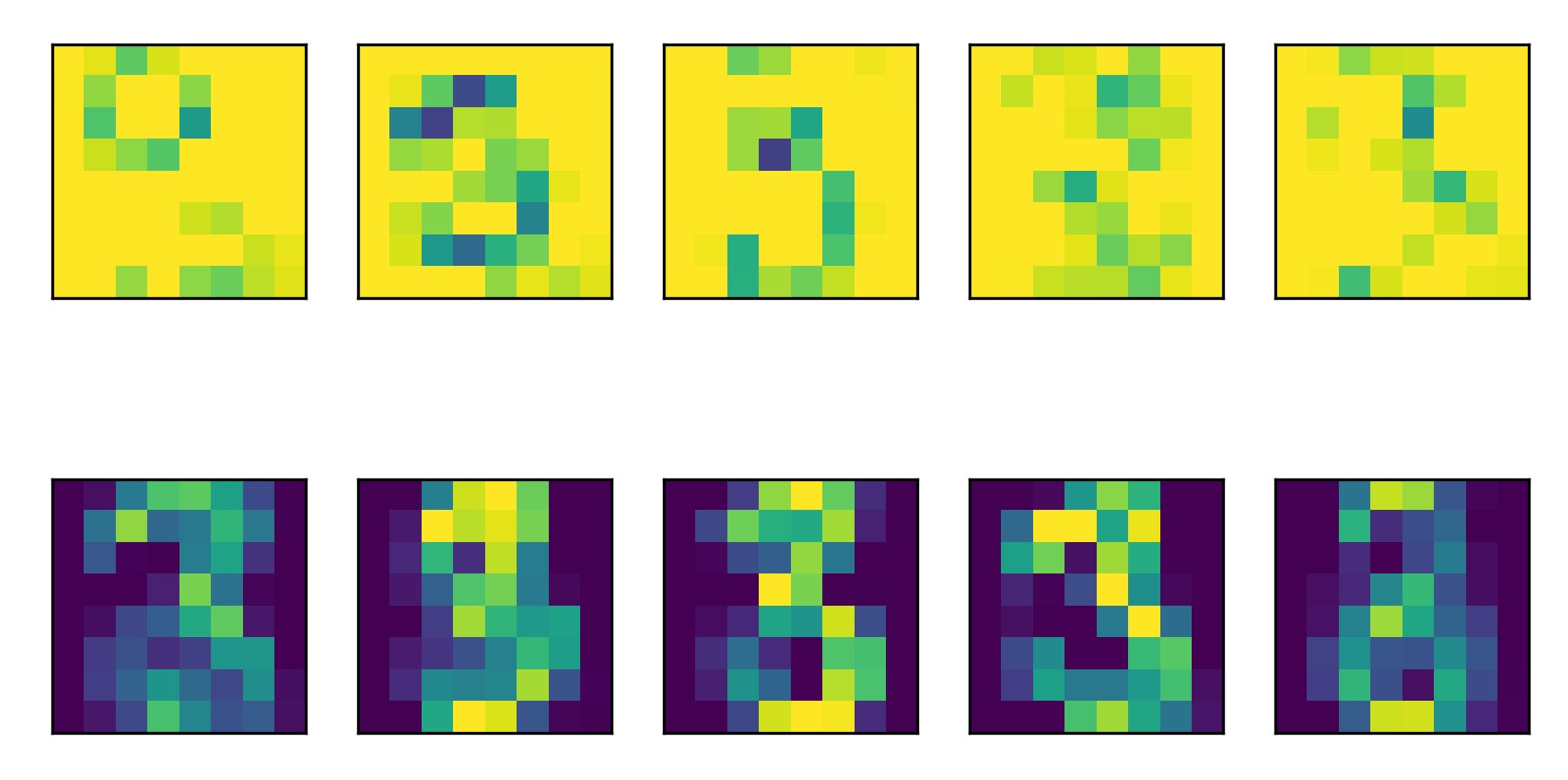}
\caption{Samples generated by the generative model ($k=30$) The top row displays the real part while the bottom row displays the imaginary part.}
\label{fig:example1_generative}
\end{figure}
\begin{figure}
\centering
\includegraphics[scale=.6]{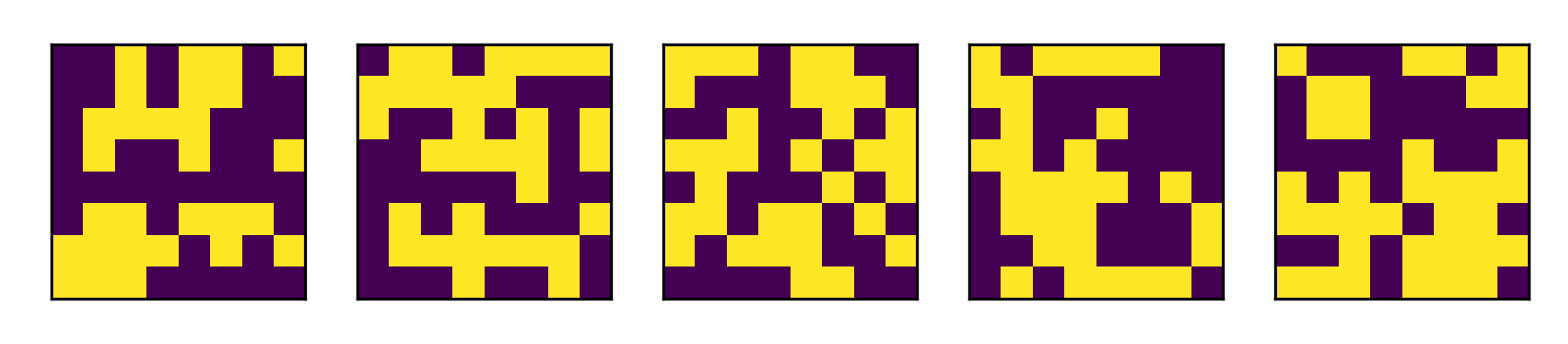}
\caption{Examples of the binary masks used to generate the measurements.}
\label{fig:example1_masks}
\end{figure}

Examples of the elements of the data set on which the generative model $\mathcal{G}$ is trained are shown in Figure~\ref{fig:example1_data}. Figure~\ref{fig:example1_generative} shows examples of the signals $\mathcal{G}({\bf z})$, ${{\bf z}\in \mathbb{C}^{30}}$, obtained from the trained generative model.  To define the measurement map $\mathcal{A}$, we use $\ell=100$ randomly generated binary probes ${\bf a_j}$, $j\in [\ell]$. Examples of the probes are shown in Figure~\ref{fig:example1_masks}. For the numerical experiment, we generate measurements with additive Gaussian noise $\boldsymbol{\varepsilon}\sim \mathcal{N}(0, \sigma^2 I_m)$. 

\begin{figure}
\begin{tabular}{cc}
\includegraphics[scale=.365]{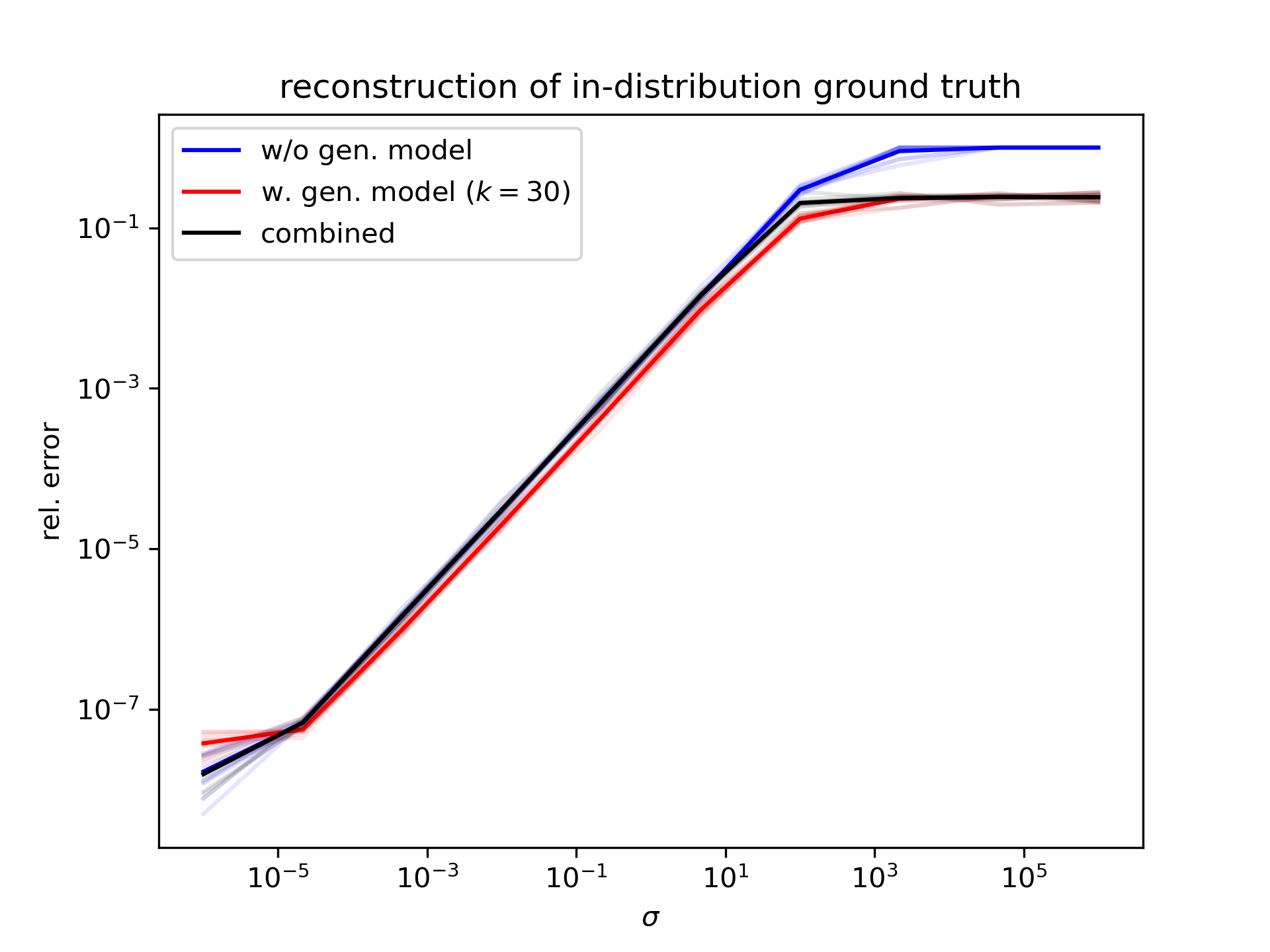} &
\includegraphics[scale=.365]{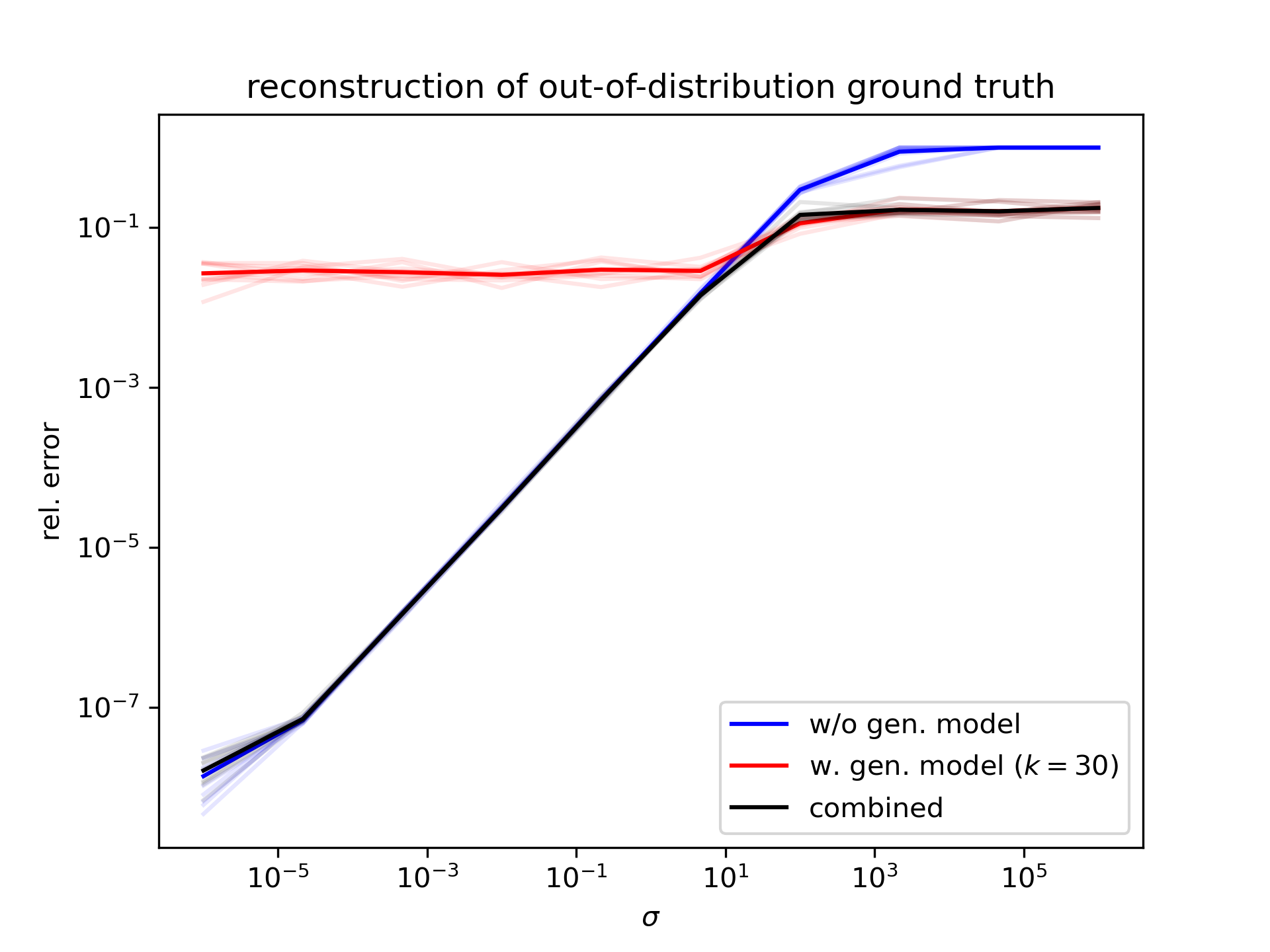}
\end{tabular}
\caption{Relative reconstruction error for the three methods for varying signal-to-noise ratio levels, tested on both in-distribution and out-of-distribution scenarios.}
\label{fig:example1_recon}
\end{figure}

To make the reconstruction methods~\eqref{eq: conventional} and~\eqref{eq: generative} more stable to the measurement noise, we introduced additional Tikhonov regularization term by setting $\lambda = \sigma^2$ and ${\bf w} = {\bf 1}$ in~\eqref{eq: unified formulation}. For the reconstruction method~\eqref{eq:combined}, we set $\lambda = 10\cdot\sigma^2$ and  ${\bf w}={\bf 1}$, which also introduces additional Tikhonov regularization term for ${\bf z}$. The reconstruction errors for the three methods, tested on both in-distribution (that is, the ground truth is generated by the generative model) and out-of-distribution (where we use samples from the original data set as ground truth) data are shown in Figure~\ref{fig:example1_recon}.

We see that for high signal-to-noise ratio levels, all methods perform well on in-distribution data, and that for low signal-to-noise ratio levels the generative model shows a slight advantage. On out-of-distribution data, the generative approach~\eqref{eq: generative} clearly shows the bias in the error for high signal-to-noise ratio regime. The combined method~\eqref{eq:combined} achieves the best result for both low and high signal-to-noise ratio levels.

\subsection{Conclusion and discussion}
\label{conclusion}
In this note, we explored the use of generative models to regularize certain inverse problems, such as phase retrieval. These preliminary results we showed indicate that generative priors can indeed improve the robustness of the inverse problem solution to measurement noise, at the expense of introducing a bias in the reconstruction. To mitigate this issue, we propose a method that aims to combine the best characteristics of both conventional and regularized methods by interpolating between them. Numerical results on phase retrieval from masked Fourier measurements show that the combined method can indeed achieve the best results.
However, the presented error bounds are rather crude and can probably be improved with more careful analysis methods. Further research is needed to solidify our understanding of the combined method, to refine it, and to make it feasible for high-dimensional problems.
In particular, the following problems should be addressed.
\begin{problem}
Improve the theoretical guarantees on the reconstruction error for a unified approach~\eqref{eq:combined}  under appropriate assumptions on the measurement map $\mathcal{A}$, generative model $\mathcal{G}$, and the ground truth ${\bf f_0}$. What if $\mathcal{A}$ does not have a (global) bi-Lipschitz property? How the bounds in Lemmas~\ref{lem: generative bound}~and~\ref{lem: unified bound} can be improved for specific classes of learned and model-based generative models~$\mathcal{G}$?
\end{problem}
\begin{problem}
Devise an efficient algorithm for solving \eqref{eq: unified formulation} with provable  convergence guarantees. Study its robustness under different assumptions on the noise $\boldsymbol{\varepsilon}$ and bias $\mathcal{G}({\bf z_0}) - {\bf f_0}$. 
In particular, as the signal-to-noise ratio level is not known in advance, the regularization parameter $\lambda$ needs to be adjusted throughout iterations of the reconstruction algorithm.
\end{problem}

\section[Structured Random Phase Retrieval]{Problem: Structured Random Phase Retrieval}
\label{S:StructuredRandomPhaseRetrieval}

\emph{Authors: Sjoerd Dirksen, Jonathan Dong,
        Felix Krahmer, 
        Palina Salanevich.}
\vspace{0.3cm}

In this section, we consider the probabilistic formulation of the noiseless phase retrieval problem. More specifically, we make the following assumptions on the object and the measurement map.
\begin{itemize}
    \item The object ${\bf f}\in \mathbb{C}^n$ is assumed to be a random vector uniformly distributed over the complex unit sphere, that is, ${\bf f}\sim \text{Unif} \left( \mathbb{S}^{n-1}\right)$.
    \item The measurement map is given by $\mathcal{A}({\bf f}) = \vert A {\bf f}\vert^2$, where $A\in\mathbb{C}^{m\times n}$ is a random matrix. The reconstruction guarantees will be drawn depending on its statistical properties.
    \item Finally, we will consider an asymptotic regime, when $m,n\to \infty$, while the oversampling ratio $\alpha = \frac{m}{n}$ stays constant. 
\end{itemize}

For simplicity of the discussion, we will focus on the noiseless case, when we are aiming to recover (an estimate) of the ground truth object ${\bf f_0}$ from its phaseless measurements 
\begin{equation}
    \label{eq: phase retrieval definition}
    {\bf y} = \mathcal{A}(\bf f_0).
\end{equation}

\begin{remark}
This phase retrieval problem formulation~\eqref{eq: phase retrieval definition} belongs to a wide class of generalized linear estimation problems. In these problems, the measurements are assumed to have the form 
\begin{equation*}
    {\bf y} = \varphi_{\text{out}}(A{\bf f_0}),
\end{equation*}
\noindent where $A\in\mathbb{C}^{m\times n}$ is a random measurement matrix
and $\varphi_{\text{out}}$ is a stochastic function. All the distributions are assumed to be known. Let $P_{\text{out}}(\cdot\vert {\bf z})$ be the probability density function associated to
$\varphi_{\text{out}}({\bf z})$.
We obtain the noiseless phase retrieval problem~\eqref{eq: phase retrieval definition}, by setting $P_{\text{out}}({\bf y} \vert {\bf z}) = \delta ({\bf y} - \vert{\bf z}\vert^2)$, where $\delta$ denotes the Dirac delta measure. For more details, we refer an interested reader to~~\cite{maillard2020phase}.
\end{remark} 

While larger values of the oversampling ratio $\alpha$ imply higher numbers of measurements and, thus, make the recovery of the object ${\bf f_0}$ a simpler task, the number of measurements available in practice is often limited. For this reason, we aim to understand the limitations of the phaseless reconstruction in the case when $\alpha$ is small. 

More specifically, the goal is to answer the following question for different classes of random sensing matrices $A$.
\begin{question}
    Depending on statistical properties of the random measurement matrix $A$, what is the smallest value of $\alpha$ for which the reconstructed approximation of ${\bf f_0}$ is better than a random guess? What is the smallest value of $\alpha$ for which the average reconstruction error is as small as can be achieved?
\end{question}
In order to formalize this question, we consider the following two reconstruction regimes and define the corresponding threshold values for $\alpha$.

Let ${\bf \tilde{f}}$ be the Bayes optimal estimator of ${\bf f_0}$ minimizing the mean square error for a given $\alpha$. Following~\cite{maillard2020phase}, we define
\begin{equation}\label{eq: MMSE_alpha}
    \MMSE (\alpha) = \lim_{n\to \infty} \mathbb{E}\Vert \bf{\tilde{f}} - {\bf f_0}\Vert_2^2
\end{equation}
\noindent to be the asymptotic mean squared approximation error. Here, the expectation is taken with respect to the distribution of the random sensing matrix $A$. We furthermore define the best possible asymptotic mean squared approximation error as
\begin{equation}\label{eq: MMSE_0}
    \MMSE_0 = \inf_{\alpha > 0} \MMSE (\alpha).
\end{equation}
\begin{itemize}
    \item \emph{Weak recovery.} In this regime, we are not looking to reconstruct ${\bf f_0}$ precisely, but rather to retrieve some information about it. We define the weak recovery threshold as 
    \begin{equation}\label{eq: alpha weak recovery}
    \alpha_{\text{WR}} = \inf \left\lbrace \alpha>0 \colon \MMSE(\alpha) < \mathbb{E}\Vert f_0 \Vert_2^2\right\rbrace.
    \end{equation}
Below $\alpha_{WR}$, any estimate will inevitably be as good as a random guess, whereas reconstruction can start to return an informed guess for $\alpha > \alpha_{WR}$. 
    
    \item \emph{Full recovery.} In this regime, we seek to achieve the best possible reconstruction in expectation with respect to the ground truth object distribution. We define the full recovery threshold as 
    \begin{equation}\label{eq: alpha full recovery}
    \alpha_{\text{FR}} = \inf \left\lbrace \alpha>0 \colon \MMSE(\alpha) = \MMSE_0\right\rbrace.
    \end{equation}
    Note that, depending on the rank of $A$, full recovery can be perfect, when $\MMSE_0 = 0$, or imperfect, when $\MMSE_0 > 0$. While full reconstruction is theoretically possible for any values of $\alpha > \alpha_{\text{FR}}$, it is not clear it it can be obtained algorithmically. 
\end{itemize}

\begin{remark}
    We note that the notion of full recovery defined above is different from the injectivity (up to a global phase factor) of the measurement map $\mathcal{A}$. While injectivity requires that any (deterministic) object can be uniquely reconstructed from its phaseless measurements, perfect full recovery only implies that $\lim_{n\to \infty} \mathbb{E}\Vert {\bf \tilde{f}} - {\bf f_0}\Vert_2^2 = 0$. In other words, for high enough input dimension $n$, the reconstruction error for almost all signals ${\bf f_0}\in \mathbb{S}^{n-1}$ is close to zero. Thus, the notion of full recovery is much weaker than the notion of injectivity. Since it is known that for injectivity of $\mathcal{A}$ it is required to have $\alpha \ge 4$~\cite{heinosaari2013quantum, fickus2014phase}, it is expected that $\alpha_{\text{FR}} \leq 4$.
\end{remark}
\begin{remark}
Alternatively to the probabilistic viewpoint discussed above, also an algebraic viewpoint has been proposed in the literature. Rather than recoverability with high probability, this viewpoint considers the space of measurement maps and asks if the subset where injectivity fails is of measure zero (typically a low-dimensional algebraic variety). There are also two versions of injectivity in this framework \cite{Li2017Identifiability}. Strong identifiability requires that almost all measurement maps have the property that no two inputs yield the same measurements, while weak identifiability is satisfied if for a fixed input $x$, almost all measurement maps do not admit any alternative input $x'$ that yields the same measurements as $x$. While the focus of our agenda is on the probabilistic notions discussed above, it would be interesting to investigate also the corresponding questions for the algebraic formulation.
\end{remark}

\subsection{Results for a random measurement matrix}\label{sec:  random matrices}

The case when the entries of the measurement matrix $A$ are i.i.d. complex Gaussian random variables, both weak and full recovery regimes are well understood. The following results were shown in~\cite{mondelli2019fundamental, maillard2020phase}:
\begin{itemize}
    \item The weak recovery threshold satisfies $\alpha_{\text{WR}} = 1$~\cite{mondelli2019fundamental}. Furthermore, the authors also showed that for any $\alpha > 1$, weak recovery of ${\bf f_0}$ can be achieved using a spectral method.
    \item The full recovery threshold satisfies $\alpha_{\text{FR}} = 2$~\cite{maillard2020phase}. To obtain this result, the authors utilized an information theory approach. Furthermore, since in this case matrix $A$ is full-rank with probability $1$, for any $\alpha > 2$ perfect full recovery is theoretically guaranteed. At the same time, it is still unknown how to achieve it algorithmically when $\alpha\approx 2$. The closest (polynomial time) algorithmic full recovery threshold is achieved by the Approximate Message Passing (AMP) algorithm, which provably achieves perfect recovery for $\alpha > 2.03$. 
\end{itemize} 

\begin{figure*}
    \centering
    \includegraphics[width=\linewidth]{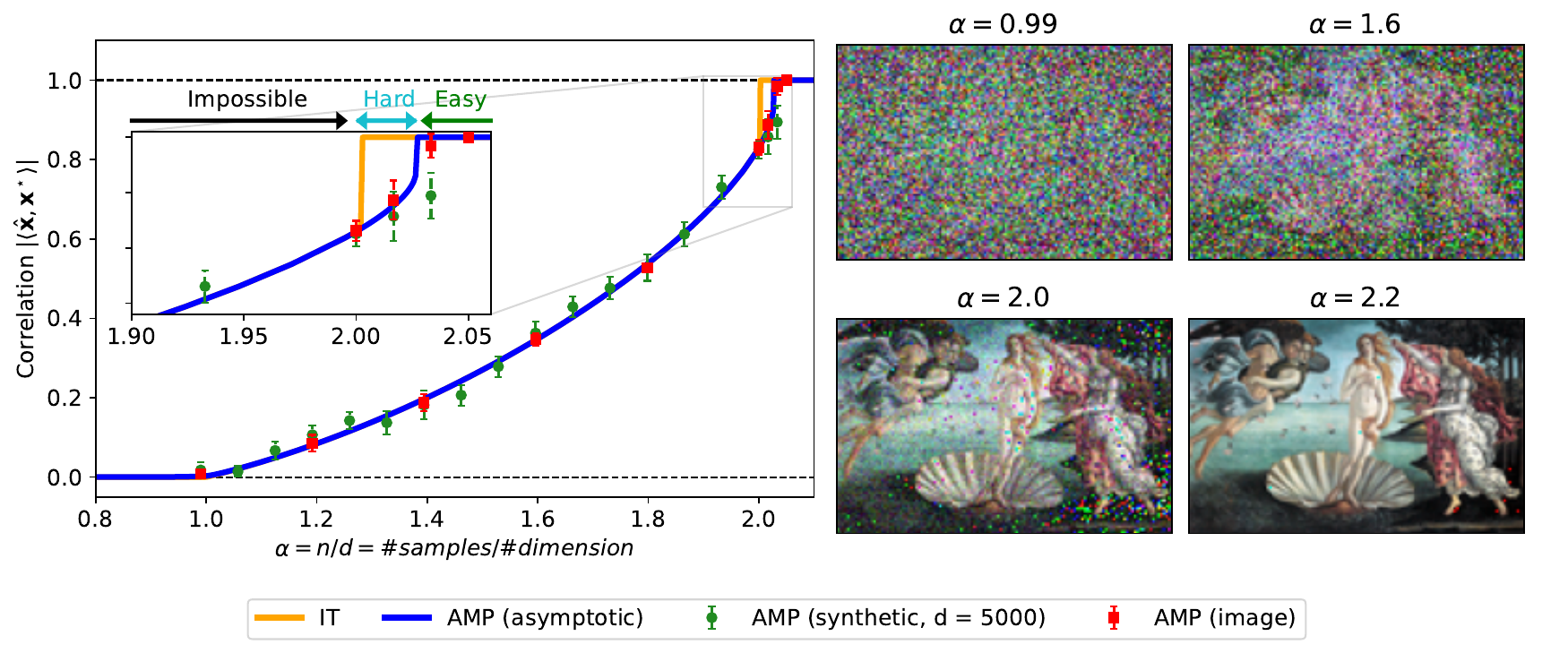}
    \caption{(Left) Correlation $| \langle \bf{\tilde f}, \bf{f_0} \rangle|$ (higher is better) achieved by the AMP algorithm in noiseless random phase retrieval as a function of the oversampling ratio $\alpha$.
    We compare it to the asymptotic theoretical prediction, known as ``state evolution'' (blue curve) and to the information-theoretic (IT) performance, which is the optimal error that any algorithm can reach, no matter its running time. 
    (Right) Examples of reconstruction from the AMP algorithm for different oversampling ratios $\alpha$.
    Figure from \cite{dong2023phase}.
    }
    \label{fig:amp}
\end{figure*} 

These results have been extended to a larger class of random measurement matrices in \cite{maillard2020phase}. Similar thresholds have been be derived for Haar-distributed (random unitary) matrices, products of i.i.d. Gaussian matrices, and right-unitarily invariant distributions. For instance, for a random unitary ensemble, both weak and perfect recovery thresholds satisfy $\alpha_{\text{WR}} = \alpha_{\text{FR}} = 2$. 

These findings establish a foundation for a robust, practical phase retrieval framework with theoretical reconstruction guarantees at the (close to) minimal number of measurements. This would be in sharp contrast with current approaches in Fourier-based or ptychographic phase retrieval settings where such theoretical guarantees are lacking. 
However, these large-scale random models are hard to physically implement, making them impractical for applications. Thus, an important problem is to construct a new class of (random) matrices that would be easily implementable while enjoying similar statistical properties. 

In practice, randomized models have been introduced in both coded diffraction imaging (CDI) \cite{candes2015phase} and random-probe ptychography \cite{pelz2017low}. These randomized physical models improve the reconstruction quality over conventional implementations \cite{pelz2017low} and reconstruction guarantees for a particular convex relaxation algorithm were first derived for $m \sim n \log^4 n$ \cite{candes2015phase}, later improved to $m \sim n \log^2 n$ \cite{gross2017improved}. This shows the gap in our understanding of these models, as for the random model, the required number of measurements $m$ scales linearly with the problem dimension $n$. 

In coded diffraction imaging and ptychography, the measurement matrix $A$ has a prescribed structure
\begin{equation}
    A=\begin{bmatrix}
        FD_{\bf w_1}\\
        \vdots \\
        FD_{\bf w_\ell}
    \end{bmatrix},
\end{equation}
where the diagonal matrices $D_{\bf w_k}$ correspond to the element-wise multiplication by probes ${\bf w_k}\in \mathbb{C}^n$, i.i.d. random independent vectors for random CDI or, shifted versions of an initial probe ${\bf w}$ in the ptychography case. Oversampling corresponds to the number of images $\ell$.



\subsection{Structured random measurement matrices}\label{sec: structured random construction}

\begin{figure}[t]
    \centering
    \includegraphics[width=0.7\linewidth]{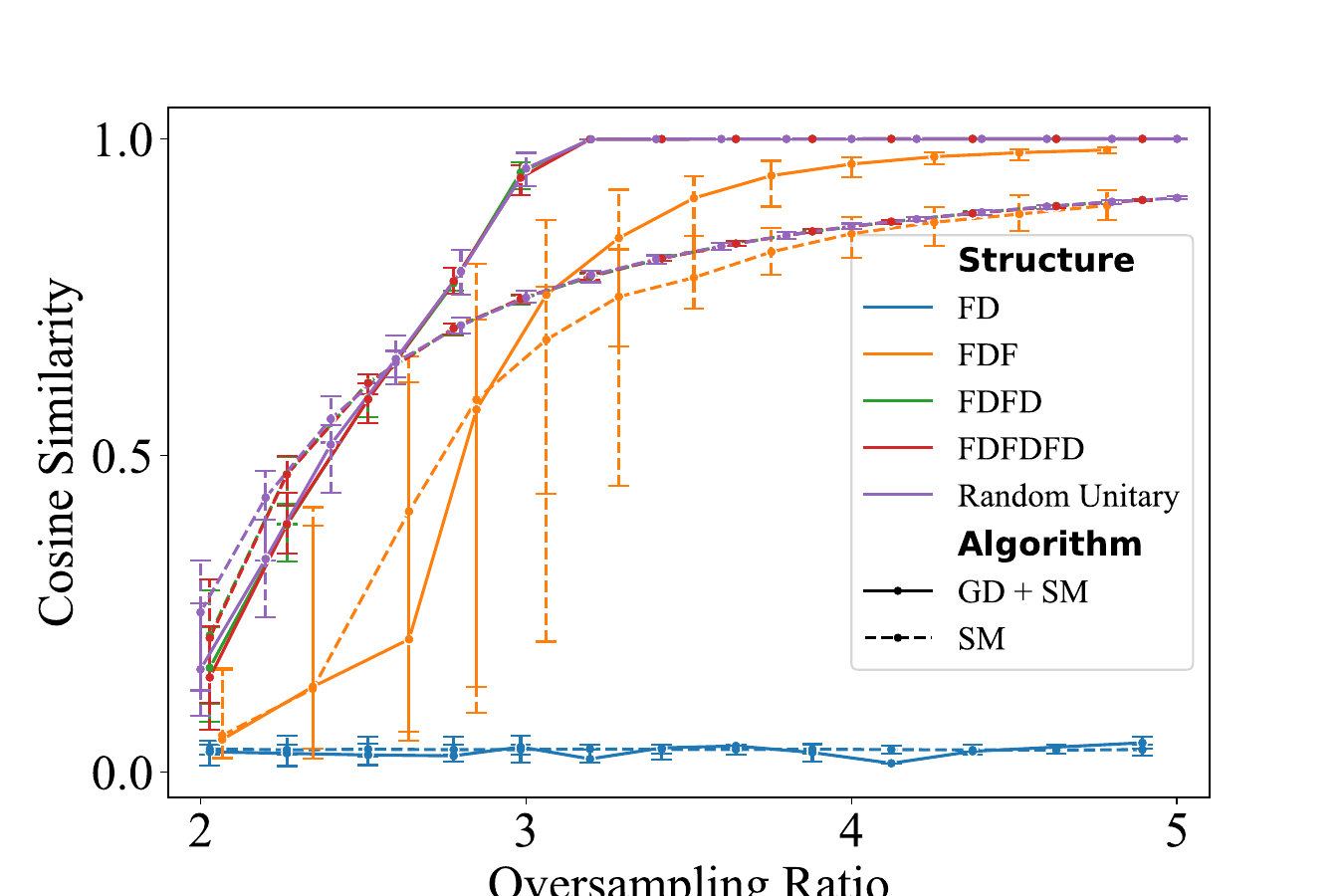}
    \includegraphics[width=0.7\linewidth]{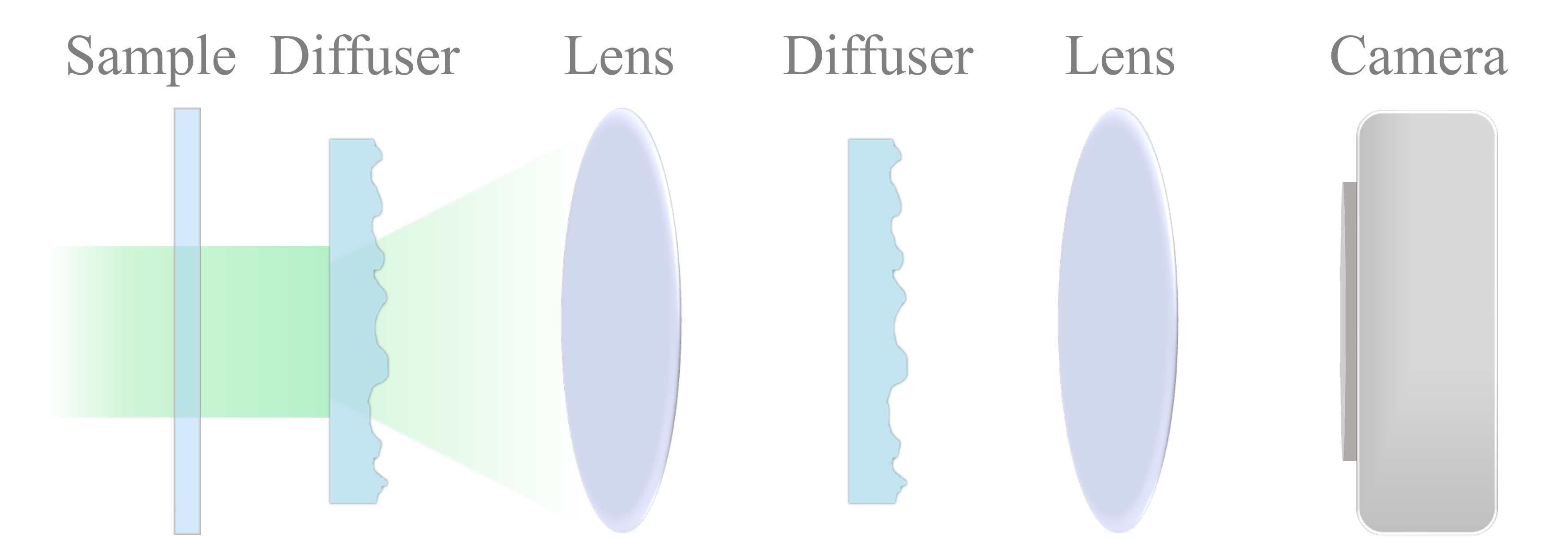}
    \caption{(Top) Performance of structured random models with several architectures and random unitary models. Two structured layers are sufficient to achieve the best results. (Bottom) Optical system implementing the structured random model. Image adapted from \cite{hu2024structured}.}
\end{figure}


To bridge the gap between realistic physical measurement models and the i.i.d. random case with strong recovery guarantees, structured random measurement matrices have recently been proposed \cite{hu2024structured}. They are built using two main optical components: optical lenses and diffusers. Lenses perform a Fourier transform $F$ on the electric field (neglecting physical scaling factors), which is similar to the operation performed by free-space propagation in ptychography and coded diffraction imaging. Optical diffusers are devices that scatter light by using a rough glass or plastic surface. Neglecting spatial correlations, they can be modeled by a random elementwise multiplication by ${\bf w}$ with entries ${\bf w}(j) = e^{\ii u_j}$ where $u_j\sim \text{i.i.d. } \text{Unif.}[0,2\pi]$ \cite{goodman2015statistical}.

To emulate the properties of i.i.d. random matrices, several lenses and diffusers can be stacked sequentially:
\begin{itemize}
\item In the measurement setup when $A = FDF$, we observe a phase transition from weak to full recovery regime after $\alpha=3$.
\item Consider $A = FD_{1}FD_{2}\cdots FD_{\ell}$ with $\ell\ge 2$ and $D_k = \diag ({\bf w_k})$, where ${\bf w_k}(j) = e^{\ii u_{k,j}}$, $u_{k,j}\sim \text{i.i.d. } \text{Unif.}[0,2\pi]$. For such measurement matrices $A$, we observe a sharp phase transition from weak to full recovery at $\alpha = 2$, suggesting the behavior similar to the Gaussian case discussed in Section~\ref{sec:  random matrices}.
\end{itemize}


The numerical findings motivate the following open problems, addressing which will lead to rigorous theoretical reconstruction guarantees for this class of physically implementable structured random measurement matrices with optimal oversampling ratio.

\begin{question}
\label{pr: threshold bounds}
    Can we derive (information theory and algorithmic) bounds on the threshold values $\alpha_{\text{WR}}$ and $\alpha_{\text{FR}}$ for the measurement matrices of the form $A = FDF$ and $A = FD_{1}FD_{2}\cdots FD_{\ell}$?
\end{question}

\medskip

In all our considerations above, we assumed that the object ${\bf f_0}$ is uniformly distributed on the complex unit sphere. However, in practice, it is often the case that there is prior information available about the ground truth object. This prior information can come in the form of positivity or support constraints, or in form of (deep) generative priors ${\bf f_0} = \mathcal{G}({\bf z_0})$ for some $z\in\mathbb{C}^d$ with $d < n$. Such prior information can make the problem of phaseless reconstruction easier, influencing the values of $\alpha_{\text{WR}}$ and $\alpha_{\text{FR}}$.

\begin{question}
    What is the dependence of $\alpha_{\text{WR}}$ and $\alpha_{\text{FR}}$ on the object priors?
\end{question}

\section[One-Bit Quantization for Event-Driven 4D STEM]{Problem: One-Bit Quantization for Event-Driven 4D STEM Acquisition in Low-Dose Applications}
\label{S:OneBitQuantization}

\emph{Authors: Wim Coene, Felix Krahmer, Allard Mosk, Palina Salanevich.}
\vspace{0.3cm}

In many imaging experiments, the measured objects, such as proteins or various cell materials, may be highly sensitive to the received dose (see, e.g.,~\cite{glaeser1971limitations}). In these cases, the measurements are performed in low-dose regime, and each detector pixel essentially measures the received illumination particle count. Due to the discrete nature of this process, the measurements at the detector pixels can be modeled by independent Poisson random variables. 

More formally, let ${\bf f}\in \mathbb{C}^n$ be the measured object, and let the phaseless measurement map be given by $\mathcal{A}({\bf f}) = \vert A{\bf f}\vert^2$, where $A\in \mathbb{C}^{m\times n}$. 
For instance, in ptychography measurement setup with probe ${\bf w}\in \mathbb{C}^n$ and $\ell$ probe positions, the measurement matrix $A$ is given by
\begin{equation}\label{eq: ptychography measurement matrix}
    A=\begin{bmatrix}
FD_1\\
\vdots \\
FD_{\ell}
\end{bmatrix}, 
\end{equation}
where $D_k = \diag (S_k{\bf w})$ and $F$ is the discrete Fourier transform matrix.
In the Poisson noise model, we assume that the entries of the measurement vector ${\bf y}$ are given by
\begin{equation*}
    y_j \sim \text{i.i.d. Poisson} (\mathcal{A}({\bf f})(j)), \qquad j\in \{1, \dots, m\}.
\end{equation*}

A commonly used approach to the phase retrieval problem with Poisson noise is negative log-likelihood minimization 
\begin{equation*}
  {\bf \tilde{f}} = \arg\min_{{\bf f}\in\mathbb{C}^n} \sum_{j = 1}^m \left( y_j\log\mathcal{A}({\bf f})(j) - \mathcal{A}({\bf f})(j) \right).
\end{equation*}
The resulting non-convex optimization problem can be solved, for instance, using ADMM~\cite{chang2018total} or Wirtinger flow algorithm~\cite{chen2015solving}. While these and similar methods show high values of the signal-to-noise ratio in the case when the values $\mathcal{A}({\bf f})(j)$ are large, their performance is limited in the extremely low dose setting~\cite{pelz2017low, li2022poisson, diederichs2024wirtinger}. 

The recent work~\cite{krahmer2024aonebit} proposes a different approach to Poisson phase retrieval under extremely low dose assumption by considering this problem from the perspective of \emph{one-bit quantization}. Indeed, already in the case when $\mathcal{A}({\bf f})(j) = 0.2$, $\mathbb{P}\left( y_j > 1\right)<0.018$, so in low-dose regime the probability of the measurements $y_j$ being outside of the set $\{0,1\}$ is very small. Furthermore, in the case of acquisition with event-driven detectors, the detectors have capacity to record only the first particle arrival, and the measurements are naturally quantized. We discuss this measurement approach in more detail in Section~\ref{sec: 4D STEM}.

To connect the considered phase retrieval problem to one-bit quantization schemes, let us consider the following. Let $\widecheck{y_j} = \min \{y_j, 1\}$, $j\in \{1,\dots m\}$, be the random quantized measurements obtained by truncating the Poisson measurements $y_j$ with threshold $1$. We have that $\widecheck{y_j}$ are independent random variables with 
\begin{equation*}
    \widecheck{y_j}\sim \text{Bernoulli} (1 - e^{-\mathcal{A}({\bf f})(j)}).
\end{equation*}
The idea explored in~\cite{krahmer2024aonebit} is to compare them to component-wise deterministic quantization of the phaseless measurements $\mathcal{A}({\bf f})$ with threshold $\delta \in (0,1)$
\begin{equation*}
    q_j = \begin{cases}
        0,& \mathcal{A}({\bf f})(j) < \delta\\
        1,& \mathcal{A}({\bf f})(j) \ge \delta
    \end{cases} = \frac{1}{2}\left(\sign (\mathcal{A}({\bf f})(j) - \delta) + 1\right).
\end{equation*}
We can now interpret the quantized measurements $\widecheck{\bf y}$ as one-bit quantization ${\bf q}$ of $\mathcal{A}({\bf f})$ corrupted by independent random bit flip noise.

For this problem reformulation, the authors in~\cite{krahmer2024aonebit} proposed a hybrid reconstruction method that uses regularized nuclear norm minimization with appropriate constraints as an initialization for the gradient descent minimization of a loss function incorporating the one-bit quantization assumption. 
In numerical experiments, where the measurement matrix $A$ is chosen to be a random complex Gaussian matrix with independent entries and the number of measurements is $m = 4n$, the proposed hybrid method shows performance that is superior to the previously studied methods~\cite{li2022poisson, diederichs2024wirtinger}. In these experiments, the low dose regime is simulated via scaling the variance of the entries of $A$. 
We note that in this measurement scenario, the number of measurements $m = 4n$ is not sufficient for full reconstruction, thus the algorithms performance is measured in terms of \emph{weak recovery}. In this regime, we aim for the relative reconstruction error (RRE) to be smaller than one, implying that the reconstructed approximation of the object is ``better than a random guess''. This is empirically achieved by the method proposed in~\cite{krahmer2024aonebit}. At the same time, it is remarkable that for zero initialization (with $\text{RRE} = 1$), the methods from~\cite{li2022poisson, diederichs2024wirtinger} converge to an approximation with $\text{RRE} > 1$, suggesting that their iterations converse to suboptimal local minima. 

To make a transition from weak to full recovery (with $\text{RRE} \ll 1$), one needs to increase the number of measurements. To avoid the damage and preserve the overall dose received by an object, this would lead to reducing the dose per pixel value. In this setting, most of the pixels will receive $0$ particles, and only a few measurements will be $1$. This measurement scenario is realized in event-driven 4D STEM methods.

\subsection{Event-driven 4D STEM acquisition}\label{sec: 4D STEM}

The imaging method we discuss in this section is known as \emph{4D-scanning transmission electron microscopy} (4D-STEM) with event driven detectors, which are originally developed at CERN for detection of elementary particles. In this method, a probe beam of high-energy electrons is scanned over the object on a 2D grid, and the transmitted electrons are detected on a 2D detector working in \emph{event mode}~\cite{jannis2022event}. That is, it detects every incident electron and outputs the time-code, as well as its $x$ and $y$ position. Such an event-driven detector can register a diffraction pattern for each of $10^{6}$ probe positions in $1$ second, and the detector consist of $256\times 256$ pixels, with about $50$ particle counts per probe position. This imaging method, among other benefits, offers high reconstruction precision and high dose efficiency, with spatial resolution below 0.05nm. This makes this technology very appealing for applications in material science, including newly emerging 2D materials.

In this measurement setup, for each probe position, we have on average $50/256^2 \approx 7.6 \times 10^{-4}$ detections per pixel of the detector, that is we operate in extremely low dose regime.  
However, close to the center of the measured diffraction patterns (where the so-called $0^{\text{th}}$ diffraction order resides), the intensity is typically a factor $10^4$ higher than in the other (possibly diffuse) diffraction orders. Hence in this area the probability of having $y_j>1$ is not negligible, but can still be assumed small (estimated to be of order $0.2$).

\begin{problem}
    Design an efficient reconstruction method for the setting of 4D STEM event-driven acquisition with $m\gg n$ and low overall dose received by the object. Prove theoretical reconstruction and stability guarantees in this measurement setup.
\end{problem}

As in this event-driven detection scenario the measurements are naturally quantized, it is particularly interesting to see if the method proposed in~\cite{krahmer2024aonebit} can be adapted to the setting with $m\gg n$, with even lower dose per measurement, and the measurement matrix $A$ given by~\eqref{eq: ptychography measurement matrix}.

\section[Mathematical Foundations of Wigner Distribution Deconvolution]{Problem: Mathematical Foundations of Wigner Distribution Deconvolution}
\label{S:MathematicalFoundationsOfWDD}

\emph{Authors:  Mark Iwen, Felix Krahmer,
        Oleh Melnyk, Matthias Wellershoff.}
\vspace{0.3cm}

One well-known algorithm for solving ptychographic reconstruction is Wigner Distribution Deconvolution (WDD) introduced by Rodenburg and Bates in \cite{Rodenburg.1992}. It is based on the following identity originally derived by Leon Cohen in a slightly more general form \cite{Cohen.1966}.
The intensity measurements\footnote{Note that these are identical to form to, e.g., \eqref{eq: measurements} up to $\Omega \times R \neq {\mathbb R}^d \times {\mathbb R}^d$.} 
\begin{equation}\label{eq: meas cont}
I({\bf x},{\boldsymbol \xi}) = | \mathcal{F}[f \cdot S_{\bf x} w] (\boldsymbol{\xi}) |^2,
\quad {\bf x},{\boldsymbol \xi} \in \mathbb R^d,
\end{equation}
with object $f$ and window $w$, can be rewritten in the form of a convolution
\begin{equation}\label{eq: meas conv}
I = \mathcal W_f * \mathcal K_w,    
\end{equation}
where $\mathcal W_f$ is the Wigner distribution of $f$ given by
\[
    \mathcal W_f({\bf x},{\boldsymbol \xi}) \coloneqq \int_{\mathbb R^d} f({\bf x} + {\boldsymbol \tau}/2) \overline{f({\bf x} - {\boldsymbol \tau}/2)} e^{- \ii \langle {\boldsymbol \xi} , {\boldsymbol \tau} \rangle } d{\boldsymbol \tau}
    \quad \text{and} \quad  
    \mathcal K_w({\bf x},{\boldsymbol \xi}) \coloneqq \mathcal W_{w}(-{\bf x},{\boldsymbol \xi}).
\]
Applying both spatial and temporal Fourier transforms separates the object term from the window term,
\begin{equation}\label{eq: wdd cont}
    \mathcal{F}^{-1} I (\boldsymbol{\xi},\boldsymbol{x}) = \mathcal{A}_f(\boldsymbol{x},-\boldsymbol{\xi}) \cdot \mathcal{A}_w(\boldsymbol{x},\boldsymbol{\xi}),
\end{equation}
where the \emph{ambiguity function} is given by 
\begin{equation}\label{eq: ambiguity function}
    \mathcal A_f({\bf x},{\boldsymbol \xi}) \coloneqq \mathcal{F}^{-1} \mathcal W_f(-{\boldsymbol \xi}, \boldsymbol{x}) = e^{\ii / 2 \langle \boldsymbol{x}, {\boldsymbol \xi} \rangle} \cdot \mathcal F [f \cdot S_{\bf x} \overline f ]({\boldsymbol \xi}).
\end{equation}
If $w$ is known and $\mathcal{A}_w \neq 0$ almost everywhere, we can recover the ambiguity function of $f$ and then reconstruct $f$ from $\mathcal A_f$. 

Since its introduction in 1992, WDD has been used both for the reconstruction from measurements \cite{Chapman.1996, Mitsuishi.2023, Yang.2017} and the analysis of uniqueness and stability of the ptychographic phase retrieval problem \cite{Alaifari.2021, Bojarovska.2016, Grohs.2024a}. A particular interest was dedicated to the discretization of the procedure, which was independently established in the acoustics \cite{Richman.1998} and imaging communities \cite{Iwen.2016, Li.2017}. 

For the discrete measurements
\[
{\bf I}_{\xi,x} = | F[{\bf f} \circ S_{x} {\bf w} ]_{\xi} |^2,
\quad x \in \mathcal X \subset [n], \quad \xi \in [n],
\]
an analogy of \eqref{eq: wdd cont} applies
\begin{equation}\label{eq: wdd disc}
(F^{-1} {\bf I} F)_{\xi,x} 
= F \left[{\bf f} \circ S_{\xi}\overline{{\bf f}}  \right]_{x} 
\cdot \overline{ F \left[\overline{{\bf w}} \circ S_{\xi} {\bf w}  \right] }_{x},  
\end{equation}
if all scan positions of the window are provided, i.e., $\mathcal X = [n]$ \cite{Cordor.2020, Perlmutter.2020}. 

If the window is localized, i.e., $\operatorname{supp}(w) = [\delta]$ for some $\delta < n/2$ and we assume that 
\begin{equation}\label{eq: conditioning}
F \left[\overline{{\bf w}} \circ S_{\xi} {\bf w}  \right]_{x} \neq 0 
\quad \text{for} \quad x \in [d], \quad \xi \in [\delta],    
\end{equation}
then all Fourier coefficients $F \left[{\bf f} \circ S_{\xi}\overline{{\bf f}}  \right]_{x}$, $x \in [d]$, $\xi \in [\delta]$, and respective vectors ${\bf f} \circ S_{\xi}\overline{{\bf f}}$, $\xi \in [\delta]$ are recovered in $\mathcal O(n \delta \log n)$ operations.

\subsection{Vanishing ambiguity function and Subspace Completion}

In stark contrast to the continuous case, where vanishing of the ambiguity function $\mathcal{F}^{-1} \mathcal K_w$ is allowed on a set of measure zero, \eqref{eq: conditioning} does not allow any zeros. This is particularly problematic as it excludes symmetric windows persistent in applications \cite{Forstner.2020}. Therefore, a natural question arises:

\begin{question}\label{q: zero singular}
How many zero coefficients $F \left[\overline{{\bf w}} \circ S_{\xi} {\bf w}  \right]_{x}$ are allowed so that ${\bf f}$ may be recovered uniquely?
\end{question}

Recent results provide an example with a single zero that allows for unique reconstruction \cite{Bartusel.2023}, while losing all zero-frequency components $F \left[\overline{{\bf w}} \circ S_{\xi} {\bf w}  \right]_{0}$, $\xi \in [\delta]$, leads to ambiguities \cite{Melnyk.2024}. As for symmetric windows, they fall in between the two cases above by having at least one zero coefficient in every second vector ${\bf f} \circ S_{\xi}\overline{{\bf f}}$, $\xi \in [\delta]$. 

Since $F[{\bf f} \circ S_{\xi}\overline{{\bf f}}]$ are all constructed from the same vector ${\bf f}$, there is a redundancy in the number of the coefficients and recovery of $n$ unknowns should be still possible even if a few out of $n \delta$ coefficients are lost. 

By viewing ${\bf f} \circ S_{\xi}\overline{{\bf f}}$ as diagonals of rank-one matrix ${\bf f} \, {\bf f}^*$, we can link the Fourier coefficients $F[{\bf f} \circ S_{\xi}\overline{{\bf f}}]_x$ to one another. One possibility is to use the low-rank structure of ${\bf f} \, {\bf f}^*$ to formulate recovery as a low-rank recovery problem \cite{Lee.2016}. In \cite{Forstner.2020}, this relation is used in the case when one diagonal ${\bf f} \circ S_{\xi_0}\overline{{\bf f}}$ for some $\xi_0 \in [\delta]$ is completely recovered and the rest of the diagonals are missing at most one Fourier coefficient, similarly to the symmetric window case. Then, the missing Fourier coefficients are estimated by solving a specially constructed linear system. Although it is yet unclear, whether this procedure, named Subspace Completion (SC), theoretically allows for a unique recovery, its empirical performance suggests so.

The main drawback of SC is that its theoretical underpinning is limited to date to the specific case where at most one coefficient is lost, without a clear way to extend it for scenarios with more missing coefficients. Such settings arise, when the contribution of the Fourier coefficients $F[{\bf f} \circ S_{\xi}\overline{{\bf f}}]$ falls below the noise level. Then, 
\eqref{eq: wdd disc} can be regularized by truncation for better noise robustness, i.e., all coefficients $F[{\bf f} \circ S_{\xi}\overline{{\bf f}}]_x$ are set to zero whenever $|F \left[\overline{{\bf w}} \circ S_{\xi} {\bf w}  \right]_{x}|$ is small. As typically $|F \left[\overline{{\bf w}} \circ S_{\xi} {\bf w}  \right]_{x}|$ are small for $x$ and $\xi$ close to $n/2$, this procedure is similar to applying a low-pass filter. 

In turn, the equation
\[
F \left[{\bf f} \circ S_{\xi}\overline{{\bf f}}  \right]_{x} = \tfrac{1}{n} e^{-\frac{2 \pi i x \xi}{n}} F \left[ F{\bf f} \circ S_{-x} \overline{F {\bf f}}  \right]_{\xi},
\]
derived in \cite[Lemma 3]{Perlmutter.2020}, links bandlimitedness of the object and diagonals of its rank-one representation. This suggests that the uniqueness of reconstruction could be achieved for band-limited objects even if some Fourier coefficients of diagonals are not recovered via \eqref{eq: wdd disc}. 

The following three open questions summarize these considerations. 
\begin{question}\label{q: more singular}
Can a version of SC be designed to handle an arbitrary number of missing coefficients with a reasonable operational cost? 
\end{question}
\begin{question}\label{q: sc theory}
Can the performance of SC be theoretically justified?  
\end{question}
\begin{question}\label{q: truncation}
Which band limit has to be assumed to guarantee unique recovery for symmetric windows and, more generally, what is the quantitative relation between vanishing $|F \left[\overline{{\bf w}} \circ S_{\xi} {\bf w}  \right]_{x}|$ and band-limited ${\bf f}$ in terms of WDD?  
\end{question}

\subsection{Scanning Patterns in the Discrete Setting}

As mentioned before, $\mathcal X = [n]$ is a crucial assumption for \eqref{eq: wdd disc} to hold. However, all possible scans are typically not available in practice and \eqref{eq: wdd disc} has to be adapted to other scanning patterns. The only scenario, in which a version of \eqref{eq: wdd disc} is established so far, is when the scan positions from a grid, $\mathcal X = \{ sk : k \in [n/s] \}$. Then, bandlimited \cite{Perlmutter.2020,} or piecewise constant \cite{Melnyk.2023b} objects can be recovered via \eqref{eq: wdd disc}. 

At the same time, working with grid scans is undesirable due to the ambiguities arising in blind ptychography, when the window is unknown \cite{Bendory.2020} and, instead, spiral scanning patterns \cite{Huang.2014} or randomly perturbed grids \cite{Fannjiang.2020b} are used. This leads to the following question. 

\begin{question}\label{q: scan position}
Can WDD be derived for arbitrary (reasonable) scanning patterns and which object assumptions are necessary?
\end{question}

We conjecture that the answer is positive for two reasons. First of all, equation \eqref{eq: wdd disc} can be viewed as a discretization of \eqref{eq: wdd cont} and, therefore, with a proper quadrature rule, one should be able to derive \eqref{eq: wdd disc} for arbitrary $\mathcal X$. Secondly, it seems possible to apply deconvolution techniques for partially known measurements.

Scanning patterns can also be investigated for the continuous problem \eqref{eq: meas cont} with subsampled measurements $({\bf x}, \boldsymbol \xi) \in \Omega$. Due to the higher complexity of the problem, much more fundamental questions are still open, even the question of the uniqueness of the solutions is still of mathematical interest in many cases.  We will now briefly review relevant work in this setting.   


\subsection{Sampled STFT phase retrieval with compactly supported functions}

Consider the recovery of an object $f \in L^2({\mathbb R})$ from the magnitude of its \emph{short-time Fourier transform (STFT)},
\begin{equation*}
    {\mathcal V}_w f (x,\xi) := \int_{\mathbb R} f(t) w(t-x) e^{-\ii t \xi} ~d t = \mathcal{F}[f \cdot S_{x} w] (\xi), \qquad x, \xi \in {\mathbb R},
\end{equation*}
where $w \in L^2({\mathbb R})$ is the window. The STFT is also called the \emph{Gabor transform} when $w$ is chosen to be the Gaussian $t \mapsto 2^{1/4} \exp(-\pi t^2)$. In this case, we denote ${\mathcal G} := {\mathcal V}_w$.

In \emph{sampled STFT phase retrieval}, one is interested in the recovery of $f$ from $|{{\mathcal V}_w f}|^2_\Omega := \{ I(x,\xi) ~|~ (x,\xi) \in \Omega\}$ with $I$ as per \eqref{eq: meas cont} for $d=1$, and $\Omega \subset {\mathbb R}^2$ a, e.g., countable \emph{lattice}, \begin{equation*}
    \Omega := \{ k {\bf v}_1 + \ell {\bf v}_2 ~|~ k,\ell \in {\mathbb Z} \} \subset {\mathbb R}^2,
\end{equation*}
for a basis $\lbrace {\bf v}_1, {\bf v}_2 \rbrace$ of ${\mathbb R}^2$.  Note that the forward operator $f \mapsto |{{\mathcal V}_w f}|^2_\Omega$ is invariant under the action of 
$\mathbb{S}^0 = \{ e^{\ii \alpha} ~|~ \alpha \in {\mathbb R} \} \subset {\mathbb C}$ by scalar multiplication on $L^2({\mathbb R})$. Hence, as above, one instead attempts to recover the orbits of this action, which is referred to as recovery \emph{up to a global phase} in the literature. To describe this the equivalence relation
\begin{equation*}
    f \sim g \iff \exists \alpha \in {\mathbb R}: f = e^{\ii \alpha} g
\end{equation*}
is commonly used.

Let $B > 0$ and suppose that $f$ is supported in $[-B,B]$.  It is known that $\exists a,b > 0$ such that, for $\Omega = a \mathbb{Z} \times b \mathbb{Z}$, all sufficiently smooth and compactly supported functions $f \in L^2([-B,B])$ whose Fourier series coefficients (after periodically extending $f$ to all of $\mathbb R$) decay monotonically in an averaged sense can be recovered up to a global phase using WDD with windows $w$ that consist of the product of a trigonometric polynomial with a characteristic function whose support includes the support of $f$ as a sufficiently small subset \cite{iwen.2023, Iwen.2020}. 
In the Gabor setting it is also known that compactly supported functions $f \in L^p([-B,B])$, where $p \in [1,\infty]$, can be recovered up to a global phase from their sampled Gabor transform magnitudes on $\Omega = {\mathbb N} \times 2 \pi b {\mathbb Z}$ if $b < (4B)^{-1}$ is just slightly smaller than half the Nyquist--Shannon sampling rate \cite{Grohs.2023,Wellershof.2024}. The same applies to bandlimited functions $f$ with $\Omega = 2\pi b {\mathbb Z} \times {\mathbb N}$. We also note that the result extends to sampling on $\lbrace 0, 1 \rbrace \times \tfrac{\pi}{2B} {\mathbb Z}$ when $p \geq 2$ \cite{Wellershof.2023}. So, only two time bands are actually necessary for uniqueness up to a global phase in the Gabor setting.  

\begin{question}
    What happens when $w$ is not, e.g., the Gaussian or a simple window of the type utilized in \cite{iwen.2023,Iwen.2020}? What other window functions $w \in L^2({\mathbb R})$ allow recovery up to a global phase from STFT magnitude measurements sampled on a lattice $\Omega$ (cf.~also \cite{Grohs2022From})?
\end{question}

\begin{question}\label{q: scan position cont}
Fix any sufficiently ``nice" window function $w$.  Do uniqueness results hold for spiral-like sampling sets $\Omega \subset {\mathbb R}^2$?  More generally, what exactly determines when a given sampling set $\Omega$ is sufficiently informative to allow for recovery of $f$ up to a global phase from $|{{\mathcal V}_w f}|^2_\Omega$? 
\end{question}

A natural question related to the paragraph above is whether it is possible to remove the assumption that our signals $f \in L^2({\mathbb R})$ are compactly supported (or bandlimited). More precisely, can we recover any given $f \in L^2({\mathbb R})$ up to a global phase from $|{{\mathcal V}_w f}|^2_\Omega$?  And, if so, with what window function(s) $w$ and sampling set(s) $\Omega \subset {\mathbb R}^2$?  A related result \cite{Alaifari.2022} shows that $f \in L^2({\mathbb R})$ cannot be recovered up to a global phase from $|{{\mathcal G} f}|^2_\Omega$ measured on any set $\Omega$ of infinitely many equidistant parallel lines in ${\mathbb R}^2$, and this remains true for all other windows $w \in L^2(\mathbb{R})$ as well \cite{grohs2022foundational}. Therefore, the recovery of all $f \in L^2({\mathbb R})$ up to a global phase is impossible when $\Omega$ is any lattice (since lattices are always contained in such a set of equidistant parallel lines).

\begin{question}
A more tractable problem than considering the recovery of all $f \in L^2({\mathbb R})$ up to a global phase using {\it some} window function $w$ might be to instead focus on related problems in the Gabor setting.  For instance, there is a dense set of functions in $L^2({\mathbb R})$ that \emph{cannot} be recovered up to a global phase from $|{{\mathcal G} f}|^2_{a {\mathbb Z} \times b {\mathbb Z}}$ while the Gaussian $w$ \emph{can} be recovered up to a global phase from $|{{\mathcal G} w}|^2_{a {\mathbb Z} \times b {\mathbb Z}}$ \cite{alaifari2024connection}. Is there a dense set of other functions that can also be recovered up to a global phase from Gabor magnitude measurements on $\Omega = a {\mathbb Z} \times b {\mathbb Z}$? If so, is this set of the first or second category?  This final question arises from the desire to understand whether we are more ``likely'' to encounter a function that can be recovered from Gabor magnitude measurements or one that cannot.
\end{question}

In summary, the known results concerning the recovery of $f$ from a {\it countably infinite} set of measurements $|{{\mathcal V}_w f}|^2_\Omega$ are currently of the following three types:
    \begin{enumerate}
        \item $f \in L^2({\mathbb R})$ can be recovered up to a global phase from measurements $\bigcup_{i = 1}^4|{{\mathcal V}_{w_i} f}|^2_\Omega$ for sufficiently fine lattices $\Omega$ provided that the four windows $w_i$ are suitable linear combinations of the Gaussian and the first Hermite function $t \mapsto 2^{5/4} \pi t \exp(-\pi t^2)$ \cite{Grohs.2021unpublished},
        
        \item $f \in L^2({\mathbb R})$ can be recovered up to a global phase from Gabor magnitude measurements $|{{\mathcal G} f}|^2_{\Omega}$ if $\Omega$ is a square-root lattice \cite{Grohs.2024b} or $\Omega$ is a union of three separated sets (where $\Omega$ is not separated itself) \cite{grohs2023phase}, and 
        
        \item as previously mentioned, compactly supported functions $f$ (as well as bandlimited functions) can be recovered from measurements $|{{\mathcal V}_w f}|^2_\Omega$ with special windows $w$ on sufficiently fine lattices $\Omega$ \cite{Grohs.2023,iwen.2023,Iwen.2020,Wellershof.2024,Wellershof.2023}.
    \end{enumerate}
    
Of all of these results, only \cite{iwen.2023,Iwen.2020} also considers the {\it approximate} recovery of $f$ up to a global phase from a {\it finite} set $\Omega$ of (lattice) measurements $|{{\mathcal V}_w f}|^2_\Omega$.  Indeed, such results provide a bridge linking the problems discussed in this section to the discrete WDD algorithms and theory discussed above.  As a result, the development of more results along these lines is encouraged.

\begin{question}
Develop practical phase retrieval algorithms $\mathcal{A}: [0,\infty)^{k} \rightarrow L^2({\mathbb R}^d)$ together with window functions $w$ and finite sampling schemes $\Omega_k = \{ (x_j,\xi_j)~|~ 1 \leq j \leq k\} \subset {\mathbb R}^d \times {\mathbb R}^d$ such that, e.g, 
$$\inf_{\alpha \in S^1} \| \alpha f - \mathcal{A}\left( |{{\mathcal V}_w f}|^2_{\Omega_k} \right) \|_{L^2} \rightarrow 0$$
as $k \rightarrow \infty$ for all $f$ in a dense set of functions of $L^2({\mathbb R}^d)$.  We note here that the case $d > 1$ is of particular interest given that \cite{iwen.2023,Iwen.2020} focus on $d = 1$.
\end{question}


\section*{Acknowledgments}
The Phase Retrieval in Mathematics and Applications (PRiMA) workshop, Aug. 5 -- 9 2024, which resulted in this manuscript was supported by a Lorentz Center conference grant (funded by Leiden University \& Dutch Research Council).  In addition, the authors would also like to acknowledge the following additional support:  Julien Flamant acknowledges the support of the French Agence Nationale de la Recherche (ANR), under grant ANR-23-CE48-0007 (project ATEMPORAL).  Mark Iwen was supported in part by the US National Science Foundation grants NSF DMS 2108479 and NSF EDU DGE 2152014.
Palina Salanevich is supported by NWO Talent program Veni ENW grant, file number VI.Veni.212.176. The authors would also like to acknowledge Mahdi Soltanolkotabi, Rima Alaifari, Francesca Bartolucci, and Mitchell Taylor for their valuable insights and discussions that contributed to the development of this manuscript.

\section*{Declarations}
The authors have no competing interests to declare that are relevant to the content of this article.


\bibliographystyle{plain}
\bibliography{main}

\end{document}